\documentclass[english,aps,prl, twocolumn,longbibliography]{revtex4-2}
\usepackage[utf8]{inputenc}
\usepackage{amsmath}
\usepackage{amssymb}
\usepackage{xspace}
\usepackage{wasysym}
\usepackage{bm}
\usepackage{graphicx}
\usepackage{hyperref}
\usepackage{braket}
\usepackage{babel}
\usepackage{xcolor}
\usepackage{mathtools}
\usepackage{xspace}
\usepackage{xcolor}
\usepackage{physics}
\usepackage{bbm}
\usepackage{bm}
\usepackage{times}
\usepackage{ulem}
\usepackage{xurl}
\usepackage{xcolor}
\newcommand{\ee}{\mathrm e}
\newcommand{\ii}{\mathrm i}
\newcommand{\Z}{\mathbb Z}
\newcommand{\FW}{\mathrm{FW}}

\newcommand{\Potts}{\mathrm{Potts}}
\newcommand{\Ising}{\mathrm{Ising}}
\newcommand{\clean}{\mathrm{clean}}
\newcommand{\clockmodel}{\mathrm{clock}}
\newcommand{\Ree}{\operatorname{Re}}
\newcommand{\arcosh}{\operatorname{arcosh}}
\definecolor{darkblue}{HTML}{004D6B}
\definecolor{darkred}{HTML}{8c1515}
\definecolor{darkgreen}{HTML}{006400}
\usepackage{hyperref}
\hypersetup{
    pdftitle={ThresholdReplicaTheory},
	colorlinks=true,
	urlcolor=darkred,
	citecolor=darkblue,
	linkcolor=darkred,
	breaklinks
}
\pagecolor{white}

\begin{document}

\title{Nishimori Threshold Estimation for Bayesian Inference and $\mathbb{Z}_q$ Surface Code Decoding}
\author{Rohit Mukherjee}
\affiliation{Institute for Theoretical Physics, University of Cologne, Zulpicher Straße 77, 50937 Cologne, Germany}

\author{Simon Trebst}
\affiliation{Institute for Theoretical Physics, University of Cologne, Zulpicher Straße 77, 50937 Cologne, Germany}

\date{\today}


\begin{abstract}
In quantum error correction, the error threshold provides essential quantitative guidance for
the ability to bring about fault-tolerance through decoding the effects of incoherent noise, 
weak measurement or inference. However, the numerical value of an error threshold is
typically only accessible through large-scale numerical simulations of the underlying noise model.
Here we introduce an analytical estimate of error thresholds falling into the Nishimori universality class
via a Fourier--Walsh projection scheme that maps the critical point of the underlying
disorder-free statistical-mechanics model to the Born-disordered Nishimori critical point.
Using a minimal replica theory approach, this closed-form estimate is obtained from a projection 
of the exact replicated single-bond weight which we find to reproduce (within a percentage point) the known numerical thresholds
of random-bond and random-plaquette Ising models / $\mathbb Z_2$ stabilizer codes in spatial dimensions $d=2-5$, and extends to Potts
variables throughout the continuous-transition regime $q\le4$.
The main application of our projection scheme is to $\mathbb Z_q$ surface codes, whose decoding
problem maps to the disordered $q$-state clock model.  For $q\ge5$
the clean clock model has \textit{two} Berezinskii--Kosterlitz--Thouless
transitions, which the projection maps to two
Nishimori temperatures that bound an intermediate information-critical
phase.  The resulting threshold values not only accurately agree with recent
decohered-$\mathbb Z_q$-toric-code numerics, but are found to satisfy
the Gilbert--Varshamov self-dual entropy relation
$\ln q \simeq H_q(T_1^\ast)+H_q(T_2^\ast),$
although no duality condition is imposed in the construction.
Our approach thereby points to a deeper connection between the clean and Born-disordered models,
while allowing for instant analytical estimates of error thresholds for a variety of stabilizer codes.
\end{abstract}


\maketitle


In the theory of critical phenomena, universality controls the exponents and amplitude ratios 
but not the bare coupling at criticality. To pinpoint the location of a phase transition one typically
relies on numerical estimates -- with a few exceptions where analytical predictions are possible
due to an additional structure such as self-duality~\cite{KramersWannier1941}, 
gauge symmetry~\cite{Elitzur1975}, or integrability~\cite{baxter1985exactly}. 
Here we turn to the universality class of {\it Nishimori transitions}~\cite{Nishimori_1980,Nishimori1981,NishimoriBook} 
-- notable for its broad applicability to decoding problems~\cite{Dennis2002},
state preparation~\cite{ZhuTantivasadakarnVishwanthTrebstVerresen,LeeJiFisherBi,Chen2025}, 
and Bayesian inference problems~\cite{Iba_1999,PutzGarrattNishimoriTrebstZhu,NahumJacobsen2025,Patil2026,WieseDasNahum2026}
-- and, employing tools from replica theory and a minimal ansatz, provide a closed-form local map that pinpoints the
Nishimori critical point $p_c$ across Ising, Potts $(2\leq q \leq4)$, and $\mathbb{Z}_q$ clock models in arbitrary spatial dimensions
\begin{equation}
	\label{eq:ThresholdEstimate}
	\beta_c^{\rm Nishimori} (\mathbb{Z}_q) = 2  \beta_c^{\rm clean} + \frac{1}{4} \ln q \,.
\end{equation}
The only input in this construction is the critical point $\beta_c^{\rm clean}$ of the underlying statistical-mechanics model 
{\it without} any disorder. When the latter is known exactly, such as for the two-dimensional (2D)
Ising model or the family of 2D Potts models, our prediction takes on a closed form. 
For instance, for the 2D Ising model, we map its critical temperature $\beta_c = \ln{(1+\sqrt{2})}/2$
to an estimate of the Nishimori critical point at
\begin{equation}
	\label{eq:ThresholdEstimate_2DIsing}
    p_c^{\rm Nishimori}=\frac{1}{1+e^{2\beta_c^{\rm Nishimori}(\mathbb{Z}_2)}}=10.82... \% \,,
\end{equation}
as we derive, in detail, below.
Comparing our predictions with existing numerical thresholds, we find that our analytical estimates
are in rather close agreement -- often exhibiting less than 1 percentage point deviation, see Table \ref{tab:thresholds}. 

There are two limits in which we expect our  
construction to reproduce {\sl exact} threshold values. 
First, we find that our local projection gives the exact value of the error threshold \cite{bethe1,bethe2,bethe3} 
for the (loop-free) Bethe lattice for any coordination number.
Second, using arguments from renormalization group theory, we expect our analytical estimates to approach
the exact threshold values of $\mathbb{Z}_q$  clock models in the limit of large $q \gg 4$. 


Exact results for such threshold estimates are scarce. 
For the 2D $\pm J$ random-bond Ising model (RBIM) 
Nishimori, Takeda, and Sasamoto conjectured 
a duality~\cite{Nishimori_1980,TakedaSasamotoNishimori2005} 
which gives a closed-form expression for the multicritical point that is distinct
from our prediction \eqref{eq:ThresholdEstimate} above.  
Their construction has been extended to Potts gauge glasses, $\mathbb{Z}_q$ models, and
pairs of mutually dual lattices~\cite{TakedaSasamotoNishimori2005,Maillard_2003,Ohzeki_2015}, 
but each application requires a
model-specific duality relation and an ansatz on which Boltzmann
factor controls the singularity. Outside the duality framework,
thresholds are accessible primarily through model-specific numerics,
which becomes prohibitive for clock variables already at moderate~$q$, where
{\sl two} close-lying BKT transitions and their slow vortex-pair dynamics
require millions of core-hours of simulation time.

\begin{table}[t]
	\caption{{\bf Projection-based threshold estimates} 
        for hypercubic lattice geometries
		compared with numerical benchmarks. 
		For Ising and Potts models, $p_c$ indicates the physical error threshold; 
		for $\mathbb{Z}_q$ clock models, the threshold is a decoherence temperature $T^*=1/J_0^*$, see  Appendix \ref{app:ZqModels}. 
		For clock models with $q \ge 5$ there are two Nishimori transitions with an intermediate (information) critical phase. 
		The error bars for the projection method are obtained by propagating the uncertainty of the critical temperature of
		the corresponding clean model. 
		When the latter is known exactly, we provide the leading digits of the projected Nishimori temperature. 
		RBIM and RPGM denote the random-bond Ising  and  random-plaquette gauge model, respectively.
        		Bold-faced thresholds indicate, to the best of our knowledge, the first analytical 
        		estimates (independent of approach) for the Nishimori transitions in the respective models.
		The references in the last column provide the numerical thresholds of the third column.
		}
\label{tab:thresholds}
\begin{ruledtabular}
\begin{tabular}{lllc}
Model & \multicolumn{1}{c}{Projection} & \multicolumn{1}{c}{Numerics} & Ref. \\
\hline
\multicolumn{4}{c}{{\it Disordered Ising models}  [\%] }\\
\hline
2D RBIM			& $p_c=10.82...$ 				& $10.92212(4)$ 		& \cite{b8y5-k3y6}\\
3D RBIM 			& $p_c=22.6260332(14)$ 			& $23.180(4)$ 			&\cite{PhysRevB.76.184202} \\
3D RPGM 		& $\bm{p_c}\mathbf{=\phantom{0}3.2539301(6)}$ 	& $\phantom{0}3.3(1)$	&\cite{OHNO2004462} \\
4D RBIM 			& $p_c=28.220653(13)$ 			& $28.1(1)$ 			& \cite{PhysRevB.64.224430} \\
4D RPGM			& $p_c=10.82...$ 				& $10.92212(4)$ 		& \cite{b8y5-k3y6}\\
5D RBIM			& $\bm{p_c=31.41719(4)}$ 			& $32.0(10)$			& \cite{PhysRevB.44.652}\\
\hline
\multicolumn{4}{c}{{\it Disordered $q$-state Potts models in} 2D [\%] } \\
\hline
$q=3$ 			& $p_c=15.5709...$ 				& $15.6(9)$ 		& \cite{Andrist2015} \\
$q=4$ 			& $p_c=19.2981...$ 				& $18.3(13)$ 		& \cite{Andrist2015} \\
\hline
\multicolumn{4}{c}{{\it Disordered $\mathbb{Z}_q$ clock models in} 2D} \\
\hline
$q=2$ 			& $T^*=0.948...$ 					& $0.95$ 			& \cite{VijayLee2025}\\ 
$q=4$ 			& $T^*=0.474...$ 					& $0.48$			& \cite{VijayLee2025} \\
$q=5$ 			& $\bm{T_1^*=0.400099(74)}$ 				& $-$ 			&   \\
  		    		& $\bm{T_2^*=0.382990(75)}$ 				& $-$ 			& \\
$q=6$ 			& $\bm{T_1^*=0.38710(18)}$ 				& $0.38$ 			& \cite{VijayLee2025}  \\
  		    		& $\bm{T_2^*=0.30213(16)}$ 				& $0.30$ 			& \\
$q=7$ 			& $\bm{T_1^*=0.38529(18)}$ 			& $-$ 			&   \\
		    	& $\bm{T_2^*=0.23793(12)}$ 			& $-$ 			& \\                    
$q=8$ 			& $\bm{T_1^*=0.38506(18)}$ 				& $0.38$ 			& \cite{VijayLee2025} \\
  		      		& $\bm{T_2^*=0.19069(13)}$ 				& $0.19$ 			& \\
\hline
\multicolumn{4}{c}{{\it Disordered $XY$ model in} 3D/4D} \\
\hline
3D XY & $\bm{T^* =0.7869884(2)}$ & 0.7840(2) & \cite{PhysRevB.83.094203} \\
4D XY & $\bm{T^* = 1.0515000(13)}$ & $-$ &  \\
\end{tabular}
\end{ruledtabular}
\vspace{-4mm}
\end{table}

Our approach is based on a different kind of object -- a single
closed-form local map \eqref{eq:ThresholdEstimate} from the disorder strength to the clean
critical coupling that {\it simultaneously} handles Ising, Potts, and clock
variables across {\it arbitrary} spatial dimensions. The map is derived
from the exact replicated single-bond weight at a minimal replica
number, projected onto the clean order-parameter channel by a Fourier--Walsh transformation~\cite{o2021analysis,stoffer1991walsh}. 
The resulting map has a transparent two-term structure that we
derive below. It is the exact tree (cavity) criterion for the
pair-overlap channel -- the analog of the Kesten-Stigum threshold of
message-passing decoding~\cite{mossel2001informationflowtrees} -- dressed
by the minimal local loop channel, whose weight is fixed, with no
adjustable parameter, by the Fourier--Walsh projection of a single
replicated bond.
We calibrate the map's accuracy against known
Ising thresholds [see Tab.~\ref{tab:thresholds}] and algebraically identify its failure mode  for
Potts models at large $q$. 
Notably our map allows us to produce two consecutive transitions  of the {\it disordered} clock model for $q\ge 5$
from the two BKT transition temperatures of the clean clock model; in agreement with  recent Monte Carlo (MC)
results for the decohered $\mathbb{Z}_N$ toric code~\cite{VijayLee2025}. 
Our predictions for these two consecutive transitions also approximately
satisfy the Gilbert--Varshamov self-dual condition for entropy~\cite{PhysRevA.54.1098}
-- despite the projection containing no
duality input, suggesting that the construction retains a deeper
self-dual structure of the underlying clean theory.

\paragraph{Nishimori physics.---}

Nishimori criticality appears in several closely related settings. 
Originally discovered~\cite{Nishimori_1980,Nishimori1981} in the random-bond Ising model (RBIM)
it is the multicritical point $(T_N,p_N)$ 
where the ferromagnet-paramagnet boundary meets the Nishimori line
$\tanh(J \beta_N )=1-2p_N$.
This seemingly intrinsically classical physics has been connected to the quantum realm 
via several mappings starting with the problem of the toric code subject to decoherence~\cite{Dennis2002}
where the optimal decoding threshold $p_{\rm th}=p_N$ is mapped to
the critical point on the Nishimori line. 
A more recent mapping has connected the problem of state preparation using quantum 
measurements to the Nishimori line in the RBIM, identifying a threshold 
for successful state preparation \cite{ZhuTantivasadakarnVishwanthTrebstVerresen,LeeJiFisherBi,Chen2025}.
Of particular interest here will be the connection of learning transitions of the 
toric code~\cite{EcksteinPRX,PutzGarrattNishimoriTrebstZhu} 
to classical Bayesian inference settings~\cite{NahumJacobsen2025,WieseDasNahum2026}, 
for which a mapping to the Nishimori line and its critical point has previously established 
the existence of a finite learning or inference threshold~\cite{Iba_1999}. 

These examples differ in their microscopic interpretation, but they
share the same organizing principle.  On the Nishimori line the
statistical weight used for inference is matched to the probability
law that generated the disorder or measurement record.  This
Bayes-optimal matching gives the disorder-averaged theory an exact
local gauge invariance and leads to the characteristic Nishimori
identities, including the exact internal energy and relations among
gauge-invariant correlators~\cite{NishimoriBook}.  
In the language of replica theory~\cite{LeDoussalHarrisI,LeDoussalHarrisII,GeorgesHanselLeDoussalMaillard,GRL2001}
the same condition has an especially simple meaning: the planted
configuration is statistically interchangeable with a posterior sample.
Thus a theory with $R$ posterior replicas has an {\it enlarged} replica permutation 
symmetry $S_{R+1}$, and the physical Nishimori problem is obtained in the
Bayes-optimal replica limit $R\to1$.
This makes replica theory a natural starting point for a local
description of Nishimori criticality. 

\paragraph{Inference setting.---}

In order to set up our replica calculation, we turn to the Bayesian inference
setting where the Nishimori line appears in the $\beta=0$ limit which, in turn,
introduces some simplifications in our replica calculations.
The principal idea of the set up is to learn the classical configuration of an 
underlying stat-mech model (such as the Ising model) at some given 
temperature $\beta$ from probing individual bonds, i.e.\ to infer the actual 
configuration from a (noisy) picture of its domain walls~\cite{PutzGarrattNishimoriTrebstZhu,NahumJacobsen2025}.
Specifically, consider a classical configuration of spins $\sigma_j=\pm1$ on the
vertices $j$ of a square lattice with Gibbs distribution
$
P(\sigma)=e^{-\beta E(\sigma)}/\mathcal Z,
\;\;
E(\sigma)=-\sum_{\langle ij\rangle}\sigma_i\sigma_j .
$
We learn the local correlation $\sigma_i\sigma_j$ on each nearest-neighbor
bond by extracting a binary measurement outcome $m_{ij}=\pm1$ with
likelihood
$
\label{eq:binary_likelihood}
P(m_{ij}\mid\sigma_i,\sigma_j)
=
(1+\gamma m_{ij}\sigma_i\sigma_j)/2$. 
Here $\gamma\in[0,1]$ controls the measurement precision: for
$\gamma=1$ the bond variable is learned perfectly, while for
$\gamma=0$ the measurement carries no information.  The full
measurement record $\mathbf m=\{m_{ij}\}$ has a {\sl conditional} product
distribution
$
P(\mathbf m\mid\sigma)
=
\prod_{\langle ij\rangle}
(1+\gamma m_{ij}\sigma_i\sigma_j)/2.
$
After observing $\mathbf m$, Bayes' theorem gives the posterior
\begin{equation}
\label{eq:posterior}
P(\sigma\mid\mathbf m)
=
\frac{1}{\mathcal Z(\mathbf m)}
\exp\!\left[
\eta\sum_{\langle ij\rangle}m_{ij}\sigma_i\sigma_j
-\beta E(\sigma)
\right] \,,
\end{equation}
with the measurement strength reparametrized as $\gamma=\tanh\eta$.
Two temperature scales are particularly relevant. At $\beta_c^{\square}=\frac12\log(1+\sqrt2)$
the ``clean" model ($\gamma = 0$) undergoes a transition from the paramagnet to the ferromagnet. 
At $\beta=0$ the learning axis $\gamma \in [0,1]$ maps to the Nishimori line~\cite{Iba_1999},
exhibiting a transition at $\gamma_c \simeq 0.784(2)$~\cite{PutzGarrattNishimoriTrebstZhu}.
It is precisely these two transitions along the $\gamma=0$ and $\beta=0$ axes that we map onto one another 
via our replica approach.

\paragraph{Minimal-replica projection.---}
 
In setting up our replica calculation for the inference setting, we work in the $\beta=0$ limit.
Since in this limit there is no prior coupling between spins, the only local
interaction in the posterior comes from the measurement likelihood
which we use to derive the local replicated bond weight.
For $R$ replicas sharing the same measurement outcome on a bond, define
$x_a=\sigma_i^{(a)}\sigma_j^{(a)},
S=\sum_{a=1}^R x_a .$
The replicated local likelihood then is
\begin{equation}
\label{eq:replicated_likelihood_before_sum}
\prod_{a=1}^R P(m_{ij}\mid\sigma_i^{(a)},\sigma_j^{(a)})
\propto
\exp\!\left(\eta m_{ij}\sum_{a=1}^R x_a\right).
\end{equation}
For the local replicated channel weight, the same binary measurement
outcome is shared by all replicas and summed over locally.  Thus
\begin{align}
\label{eq:ising_measurement_bond_weight}
W_R(\{x_a\})
=
\sum_{m_{ij}=\pm1}
\exp\!\left(\eta m_{ij}\sum_{a=1}^R x_a\right)  
=
2\cosh(\eta S) \,.
\end{align}
So the effective replicated bond log-weight or replicated measurement averaged effective Hamiltonian is
\begin{equation}
\label{eq:ising_measurement_log_weight}
y(S)=\log\cosh(\eta S)\,.
\end{equation}
The local algebra is exact at this stage; a projection hypothesis  
enters only in the next step.  Expanding at small $\eta$ gives
$
\label{eq:logcosh_expansion}
\log\cosh(\eta S)
=
\frac{\eta^2}{2}S^2-\frac{\eta^4}{12}S^4+O(\eta^6),
$
so the leading irreducible non-Gaussian invariant is quartic. 
The number of replicas we use in Eq.~\eqref{eq:ising_measurement_log_weight} has an {\sl operational meaning} here; 
the measurement-averaged $R$-replica bond weight is the {\sl generating
polynomial} of the moments of the bond measurement outcome up to order
$R$ [cf. Eq.~\eqref{eq:WR} below], so choosing $R$ amounts to
choosing the highest local moment of the measurement record the
theory can represent. As we show below, the transition criterion on a
tree engages only second moments and is complete at $R=2$, while
$R=4$ is the smallest replica algebra containing a product of two
disjoint pair channels on the same bond, the minimal local channel
beyond the tree sector (independent of the details of the lattice geometry, see our discussion
of hexagonal lattices below). Larger $R$ retains higher local moments
independent of the lattice geometry to which they couple, and is empirically
found to over-dress the pair channel.
Finite-replica potentials have been used in glassy systems to promote
the overlap between real replicas from an observable to an order-parameter
field whose fluctuations encode the structure of the glassy
phase~\cite{finiterep1,finiterep3, finiterep2, finiterep4}. We adopt this viewpoint   
for Nishimori criticality in a {\it local} manner. The order-parameter channel is a two-replica
pair overlap $\mathcal{Q}_{ab}$: the product of two replica bonds
$\mathcal{Q}_{ab} = x_a x_b$ for Ising variables, 
the coincidence
$\delta(\tau_{ij}^{(a)},\tau_{ij}^{(b)})$ of two replica bond twists for
Potts variables, and the first harmonic of the relative bond angle for
clock variables. 
A projection onto the overlap channel can be defined with two replicas,
and at $R=2$ it is exact on trees and reproduces the Kesten--Stigum instability condition~\cite{mossel2001informationflowtrees}, as shown below. On a loopy lattice
the tree channel is no longer complete. The minimal local extension
beyond it is a product of two disjoint pair channels on the same
bond, which is first carried by the four-replica algebra.

Our minimal-replica
construction therefore starts from the exact four-replica single-bond log-weight
and projects it back onto the pair channel order-parameter  by
a Fourier--Walsh projection.
For \(R=4\), the Ising replicated log-weight
\[
  y(x)=\log\cosh\!\left(\eta\sum_{a=1}^4 x_a\right)
\]
is a function on the finite group \((\mathbb Z_2)^4\).  We expand it in the
ordinary Fourier--Walsh basis, whose characters are products
\[
  \chi_A(x)=\prod_{a\in A}x_a \,.
\]
The pair channel is spanned by the six Fourier--Walsh characters
$  \chi_{ab}(x)=x_ax_b$, with $a<b$.
By replica permutation symmetry all six pair coefficients are equal, so it is
convenient to project onto the replica-symmetric pair-overlap coordinate
\begin{equation}
\label{eq:pair_coordinate}
  X_2(x)=\sum_{a<b}x_ax_b
  =\frac{S^2-R}{2} \,.
\end{equation}
This is the unique \(S_R\)-symmetric direction in the pair-character
subspace; each individual \(x_ax_b\) is a Fourier--Walsh character, while
\(X_2\) is their symmetric sum.
The Fourier--Walsh inner product is the uniform group average,
\begin{equation}
  \langle f,g\rangle_{\rm FW}
  =
  \frac{1}{2^R}
  \sum_{x\in\{\pm1\}^R} f(x)g(x) \,.
\end{equation}
The retained pair-channel coupling is therefore the exact Fourier--Walsh
projection (see the Supplementary Material (SM) \cite{SM} for details) 
\begin{equation}
\label{eq:ising_main}
K_{\rm MRP}^{\rm Ising}(\eta)
=
\frac{\langle y,X_2\rangle_{\rm FW}}
     {\langle X_2,X_2\rangle_{\rm FW}} \,.
\end{equation}
For \(R=4\), the sum can be grouped by \(|S|\in\{0,2,4\}\).  The
multiplicities are
$W_0=\binom42=6$ , 
$W_2=2\binom41=8$, 
$W_4=2$.
Note that these are {\sl combinatorial constants} forced by
the uniform measure on \((\mathbb Z_2)^4\), not free fitting weights.
The corresponding log-weights are
\[
y_S\in
\left\{
0,\,
\log\cosh(2\eta),\,
\log\cosh(4\eta)
\right\}.
\]
Thus, the grouped Fourier--Walsh projection is
\begin{equation}
K_{\rm MRP}^{\rm Ising}(\eta)
=
\frac{\sum_S W_S X_{2,S} y_S}
     {\sum_S W_S X_{2,S}^2}
=
\frac18\log\cosh(4\eta) \,.
\end{equation}
Note that this  projected coupling separates exactly into the tree term and
the quartic-channel term
\begin{equation}
\begin{split}
  K_{\rm MRP}^{\rm Ising}(\eta)
  &= \kappa_{2}(\eta)+K_{4}(\eta) \,,\\
  \kappa_{2}(\eta)
  &= \tfrac12\log\cosh(2\eta) \,,\\
  K_{4}(\eta)
  &= \tfrac18\!\left[\log\cosh(4\eta)-4\log\cosh(2\eta)\right] \,.
\end{split}
\label{eq:treeplusloop}
\end{equation}
Here $\kappa_{2}$ is the $R=2$ pair coefficient, shown below to
reproduce the exact Bethe-lattice threshold, and $K_{4}$ is the
coefficient of the quartic character $x_{1}x_{2}x_{3}x_{4}$, the
minimal channel beyond the tree sector. It is negative for all
$\eta>0$ [with $u=\cosh2\eta$: $\cosh4\eta=2u^{2}-1<u^{4}$], so the
quartic channel always weakens the projected coupling relative to its
tree value. The decomposition makes the role of the four-replica
sector concrete: $R=4$ does not replace the pair channel, it {\it
dresses} it. In the grouped sum the $S=0$ sector is annihilated by
$y(0)=0$ and the $|S|=2$ sector by the projection factor
$(S^{2}-R)=0$ at $R=4$, so the pair projection is supported entirely
on the replica-locked sector $|S|=4$.
This coefficient is the component of the exact Nishimori replicated
single-bond log-weight along the physical pair channel.  Our projection
criterion/ansatz is that the transition is reached when this retained channel equals
the clean critical coupling
\begin{equation}
\label{eq:ising_match}
\frac18\log\cosh(4\eta_c)=\beta_c^{\rm clean}, \quad \eta_c=\tanh^{-1}{(\gamma_c)} \,.
\end{equation}
This single equation is the central technical result of our work, as it now connects the critical threshold of the Nishimori transition in closed form 
to the known critical coupling of the clean theory. 
Having established the critical coupling $\eta_c$, the inference threshold is then obtained as $\gamma_c=\tanh\eta_c$, and can be further
translated to the critical disorder strength $p_c=(1-\gamma_c)/2$ of the original RBIM, as given in Eq.~\eqref{eq:ThresholdEstimate_2DIsing}.

The minimal-replica projection~\eqref{eq:ising_match} contains a single
external input, the clean critical coupling. Inserted across the
Ising family of models (or, equivalently, the family of $\mathbb{Z}_2$ stabilizer codes), 
it can be used to pinpoint the Nishimori thresholds of both
random-bond and random-plaquette variants in spatial dimensions 
$d = 2, 3, 4, 5$, as summarized in Table~\ref{tab:thresholds}. 
As can be seen, this generality of our minimal-replica projection (MRP) scheme
allows us to
reproduce every existing MC and series-expansion threshold 
and provides the first direct analytical estimates for many disordered models (shown in bold-face in Table~\ref{tab:thresholds}).
This demonstrates that the single equation~\eqref{eq:ising_match}
captures the disorder-renormalization of the Ising channel across
the entire topological-code Ising hierarchy.

\paragraph{Exactness on the Bethe lattice.---}
A natural anchor point of our construction is the Bethe lattice, for which
the two-replica sector is already sufficient and the projection introduces
no approximation, as we show now. 
At $R=2$ the measure\-ment-averaged bond weight is itself a pure pair
coupling, summing the shared outcome
\begin{equation}
\begin{split}
  W_2(x_1,x_2)
  &=2\cosh\bigl[\eta(x_1+x_2)\bigr] \\
  &=\mathrm{const} \times \;e^{\kappa_2\,x_1x_2},
  \qquad
  \kappa_2=\tfrac12\log\cosh(2\eta) \,,
\end{split}
\label{eq:R2exact}
\end{equation}
with the useful identity $\tanh\kappa_2=\tanh^2\eta$. No projection is
involved. The two-replica algebra $\{1,\,x_1x_2\}$ is exhausted by
Eq.~\eqref{eq:R2exact}. Since $x_1x_2=Q_iQ_j$ with
$Q_i=\sigma_i^{(1)}\sigma_i^{(2)}$, the overlap spin $Q_i$ sees an ordinary
Bethe-lattice coupling $\kappa_2$, and the standard instability condition~\cite{bethe2}
$(z-1)\tanh\kappa_2=1$ becomes $\kappa_{2}=\tanh^{-1}\!\left(\frac{1}{z-1}\right)$, which can be rewritten as
\begin{equation}
  (z-1)\,\tanh^2\eta_c=1
  \quad\Longleftrightarrow\quad
  p_c=\tfrac12\Bigl(1-\tfrac{1}{\sqrt{z-1}}\Bigr) \,.
  \label{eq:KS}
\end{equation}
This is the exact threshold. Eq.~\eqref{eq:KS} is the KS
condition of the broadcasting problem (see the SM~\cite{SM} for more details and a Potts derivation), rigorously the reconstruction
threshold for the binary symmetric channel on trees, and it
coincides with the linear instability of the belief-propagation recursion
and with the tree spin-glass instability~\cite{bethe1,bethe2}. On trees the two-replica
closure is therefore a rigorous result, anchoring the construction in an exactly
solved limit.

\paragraph{Potts models.---} 
Let us now turn to the first generalization, the $q$-state Potts model with $2 \leq q \leq 4$, where
the 2D Potts model exhibits a continuous phase transition~\cite{Wu1982,duminil2017continuity}. 
For a Bayesian inference setting as in the Ising case, we calculate the local replicated bond weight for $q$-state Potts variables at $\beta=0$
\begin{equation}
	\label{eq:potts3}
	K_{\rm MRP}^{q=3 \; \rm Potts}(y)=\tfrac{1}{9}\ln\!\left[
	\frac{(y^4+2)(y^3+y+1)^2}{y^3(y+2)^3}\right],
\end{equation}
with $y=e^{J_0}$ and $J_0=\ln[(1+(q-1)\gamma)/(1-\gamma)]$. Matching
to $\beta_c(q)=\ln(1+\sqrt{q})$~\cite{Wu1982} gives $p_c\simeq 0.156$
at $q=3$. Similarly, for $q=4$ one can calculate the respective local replicated bond weight 
(detailed in Appendix \ref{app:PottsModels} and the SM~\cite{SM}) to obtain $p_c\simeq 0.193$.
These values are in good agreement (i.e.\ within error bars) with existing numerical values~\cite{Andrist2015}, 
$p_c=15.6\pm 0.9~\%$ and  $p_c=18.3\pm 1.3~\%$ , for $q=3,4$, respectively. 
For $q>4$ the clean Potts transition becomes first order~\cite{Wu1982,duminil2017continuity}. 
Our projection, however, requires the clean transition to be continuous, since a first-order transition is not characterized by a {\it single} scaling field 
and the matching condition $K_{\rm MRP}(J_0^*) = \beta_c^{\rm clean}$ no longer identifies a critical surface. 
This is a hard limitation of our construction and restricts its applicability to $q \leq 4$ Potts models. Indeed, if one naively applies our approach for $q \geq 5$ Potts models, 
one finds substantial deviations from the existing numerical estimates and Hashing bounds as shown in Fig.~\ref{fig:Potts-Hashing} of the Appendix.

\begin{figure}[t]
	\centering
	\includegraphics[width=\columnwidth]{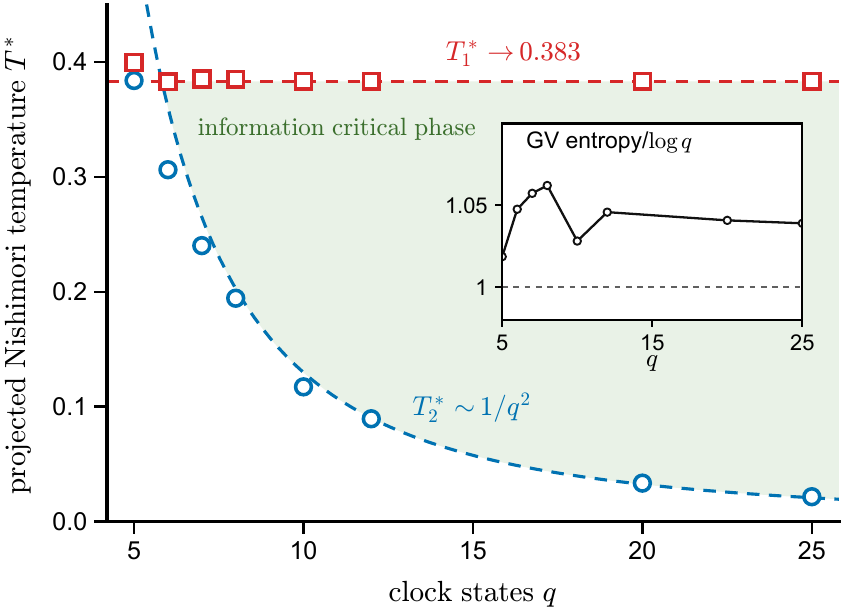}
	\caption{
		{\bf Nishimori thresholds for $\mathbb{Z}_q$ clock models.}
		Shown are the two transition temperatures $T^*_{1,2}$ of the two
		subsequent Nishimori transitions of the $\mathbb{Z}_q$ clock model
		as a function of $q$.
		While the upper temperature saturates at $T^*_1\to 0.3832$ (dotted),
		the lower projected Nishimori temperature is found to follow $T^*_2\simeq A/q^2$
		(dashed).
		The inset shows the GV self duality condition plotted as function of $q$ which, ideally, should be 1. 
		For more discussion see Appendix \ref{app:ZqModels}. 			 
		}
	\label{fig:clock_q}
\end{figure}

\paragraph{$\mathbb{Z}_q$ clock models.---} 
A natural $q$-state generalization
of the Ising model that retains a continuous phase transition for
all $q$ is the family of $\mathbb{Z}_q$ clock models, with spin
variables $\sigma_j \in \mathbb{Z}_q$ and clean nearest-neighbor
interaction $V(\sigma_i,\sigma_j) = \cos[2\pi(\sigma_i-\sigma_j)/q]$.
These models are tightly connected to Abelian quantum double models
and qudit generalizations of the toric code~\cite{Kitaev2003}.
For the Bayesian inference setting, we record a noisy bond
difference $m_{ij} \in \mathbb{Z}_q$ of the true bond
$\Delta_{ij} = \sigma_j - \sigma_i \pmod{q}$, with the discrete
von-Mises likelihood
\begin{equation}
P(m_{ij} | \sigma_i,\sigma_j) = \frac{1}{Z_q(J_0)}
\exp\!\left[J_0 \cos\frac{2\pi(m_{ij} - \Delta_{ij})}{q}\right] \,,
\end{equation}
where $J_0$ controls the measurement precision and plays the same
role as $\eta$ in the Ising case. After summing the shared
measurement outcome over $\mathbb{Z}_q$, the exact replicated
single-bond log-weight at $R=4$ depends on the occupation vector
$N = (N_0,\ldots,N_{q-1})$ with $\sum_n N_n = 4$, where $N_n$ counts
the number of replicas with bond difference $\Delta^{(a)} = n$.
The clean clock interaction is generated by the first angular
harmonic, so the natural projection coordinate is its replica
generalization, see the SM~\cite{SM} for more details
\begin{equation}
x_N = \sum_{a<b} \cos\!\left[\frac{2\pi(\Delta^{(a)}-\Delta^{(b)})}{q}\right]
= \frac{|M_N|^2 - 4}{2} \,,
\end{equation}
with $M_N = \sum_n N_n e^{2\pi i n/q}$. This is the $\mathbb{Z}_q$
analog of the Ising pair coordinate $X_2$ in Eq.~\eqref{eq:pair_coordinate} -- both project
the exact replicated log-weight onto the channel that carries the
clean order parameter. The threshold condition then becomes
$K_{\rm MRP}^{\rm clock}(J_0^*) = \beta_c^{\rm clock}(q)$ at each
clean clock critical coupling $\beta_c^{\rm clock}(q)$.


The variance decomposition of the
replicated bond log-weight provides a direct diagnostic of the
projection quality. For the clock model at the projected thresholds,
the first harmonic accounts for 94--96\% of the weighted variance of
$y_N(J_0^*)$ across $q=2,4,6,8$, 
as detailed in Table \ref{tab:harmonic-ratios} of the SM. 
Higher harmonics ($h\ge 2$) contribute at the sub-percent level; the remaining 4--6\% lies in
pure-replica directions orthogonal to the entire clock-harmonic
subspace (see the full table \ref{tab:harmonic-ratios} in the SM~\cite{SM}). The dominance of the retained
coordinate is the empirical signature that the projection captures
the leading structure of the local replicated weight, and the small
size of the higher-harmonic residual is consistent with the
empirical agreement of 
the clock model Nishimori thresholds in Table~\ref{tab:thresholds}.

For $q\ge 5$ the clean clock model has {\it two} BKT
transitions~\cite{Jose1977,Lapilli2006,Li2020,ChenHouFangDeng2022}, at
$\beta_{c1}(q)<\beta_{c2}(q)$. Inserting each clean BKT coupling
separately into the same projection equation yields two Nishimori
temperatures $T_1^*$ (upper) and $T_2^*$ (lower) as plotted in Fig.~\ref{fig:clock_q}. This is structurally
different from duality -- a single self-dual condition yields {\sl one}
predicted critical point per model, and the clean clock duality~\cite{ChenHouFangDeng2022} 
produces an
approximate self-dual point $\beta_{\rm sd}\simeq q/(2\pi)$ that lies
between the two BKT transitions without resolving them. Our MRP
construction, on the other hand, inherits the multiplicity of the clean theory and produces
{\it both} thresholds; to our knowledge no other closed-form method
previously gave both Nishimori thresholds of the disordered clock model
at finite~$q$. 
\
Although our MRP construction
contains no duality input, its predictions approximately satisfy the
Gilbert--Varshamov type self-dual condition~\cite{PhysRevA.54.1098,steane1996multiple,sasamoto2005phase} 
\begin{equation}
\label{eq:gv}
\log q= H\bigl(T_c^{(1)}\bigr)+H\bigl(T_c^{(2)}\bigr) \,,
\end{equation}
where $H(T)$ is the Shannon entropy of the von Mises distribution~\cite{VijayLee2025} 
$p_k=\exp[(1/T)\cos(2\pi k/q)]/Z(T)$. This condition has been
conjectured for disordered $\mathbb{Z}_q$ spin models on the
Nishimori line and indeed been observed for a set of small $q$ values in recent numerics~\cite{VijayLee2025}.
In our MRP approach, it emerges naturally and becomes increasingly valid
for large $q$, see Appendix~\ref{app:ZqModels}.
But the agreement is striking already for small $q=2,4$ where we find the MRP prediction 
to satisfy the conjectured self-dual condition to 1.1\%, and for $q=6,8$ it
satisfies it to within 5-6\%, as shown in  the inset of Fig.~\ref{fig:clock_q}. 
For larger $q \gg 4$, this small deviation asymptotically vanishes.
Because the MRP construction is derived from the exact replicated bond algebra 
rather than from any duality ansatz, this agreement is a nontrivial structural fact
rather than a consistency requirement. It suggests that the
Gilbert--Varshamov condition is the long-distance shadow of the
clean BKT spin-wave/vortex duality of the underlying clock model,
projected onto the Nishimori line through the first-harmonic MRP
map. The clean clock duality acts naturally on the first angular
harmonic, which is precisely the variable retained by the
projection.

\paragraph{Why $R=4$?---}
Let us return to the question of why $R=4$ is the minimal replica count
used in our projection, and why increasing the number of replicas does {\it not}
systematically improve the approximation.
A useful elementary object in this consideration is the
single-bond transmission $t_b=\tanh(\eta m_b)$; for a fixed
measurement record, this is the factor by which a two-point correlation is
multiplied upon crossing bond $b$, so that
$\langle\sigma_0\sigma_r\rangle=\prod_{b\in P}t_b$ exactly on any chain or
tree. In the language of broadcasting, this is the transfer eigenvalue of
the bond channel, whose second moment over the record enters the
KS criterion. On the Nishimori line, after
gauging the planted bond to $+1$, one has
$t=\tanh(\eta m)=m\tanh\eta=\gamma m$ with $\mathbb{E}[m]=\gamma$, hence
\begin{equation}
t^2=\gamma^2 \;\;\text{identically},\qquad
\mathbb{E}[t]=\gamma^2\,,
\label{eq:moments}
\end{equation}
where $\mathbb{E}[\,\cdot\,]$ denotes the average over the measurement
record generated by the true configuration. The equality of the first and
second moments in Eq.~\eqref{eq:moments} is the bond-level form of the
{\sl Nishimori gauge identity} \cite{Nishimori_1980}; under the disorder average, ferromagnetic and
overlap transmissions are locked, which is why the pair channel is the
natural channel carrying the clean coupling.  
As shown above, on a tree the
linear instability of the paramagnetic fixed point engages only this
two-replica transmission channel, and
$\tanh\kappa_2=\tanh^2\eta$ reproduces the KS
criterion~\eqref{eq:KS} exactly. An analogous statement holds in the
color or harmonic channel for Potts and clock variables, see the
SM~\cite{SM}. Thus $R=2$ is the complete local algebra of the tree linear
instability.

On lattices with loops, the transition is naturally monitored by
fluctuations of the Edwards--Anderson pair overlap~\cite{finiterep1}
$q_i^{(12)}=\sigma_i^{(1)}\sigma_i^{(2)}$; the lattice
counterpart of Eq.~\eqref{eq:moments},
\begin{equation}
	\mathbb{E}[\langle\sigma_0\sigma_r\rangle^2]
	=\mathbb{E}[\langle\sigma_0\sigma_r\rangle] \,,
	\label{eq:Nishimori}
\end{equation}
ties this pair-overlap
channel to the ferromagnetic channel and makes a one-channel matching
plausible. A measurement-averaged connected covariance of the overlap field
involves a disjoint replica pair,
\begin{eqnarray}
G^{\rm conn}_{ij}
&=&
\mathbb{E}\bigl[\langle q_i^{(12)}q_j^{(12)}\rangle\bigr]
-
\mathbb{E}\bigl[\langle q_i^{(12)}q_j^{(34)}\rangle\bigr] \,,
\nonumber \\[2mm]
\langle q_i^{(12)}q_j^{(34)}\rangle
&=&
\langle\sigma_i^{(1)}\sigma_i^{(2)}\rangle
\langle\sigma_j^{(3)}\sigma_j^{(4)}\rangle
\nonumber \\[2mm]
&=&
\langle\sigma_i\rangle^2\langle\sigma_j\rangle^2 \,,
\label{eq:Gconn}
\end{eqnarray}
where the last equality is understood within the fixed posterior measure.
Thus the connected two-point function of the pair overlap is naturally a
{\sl four-replica} object.

\begin{figure}[t]
	\centering
	\includegraphics[width=1.0\linewidth]{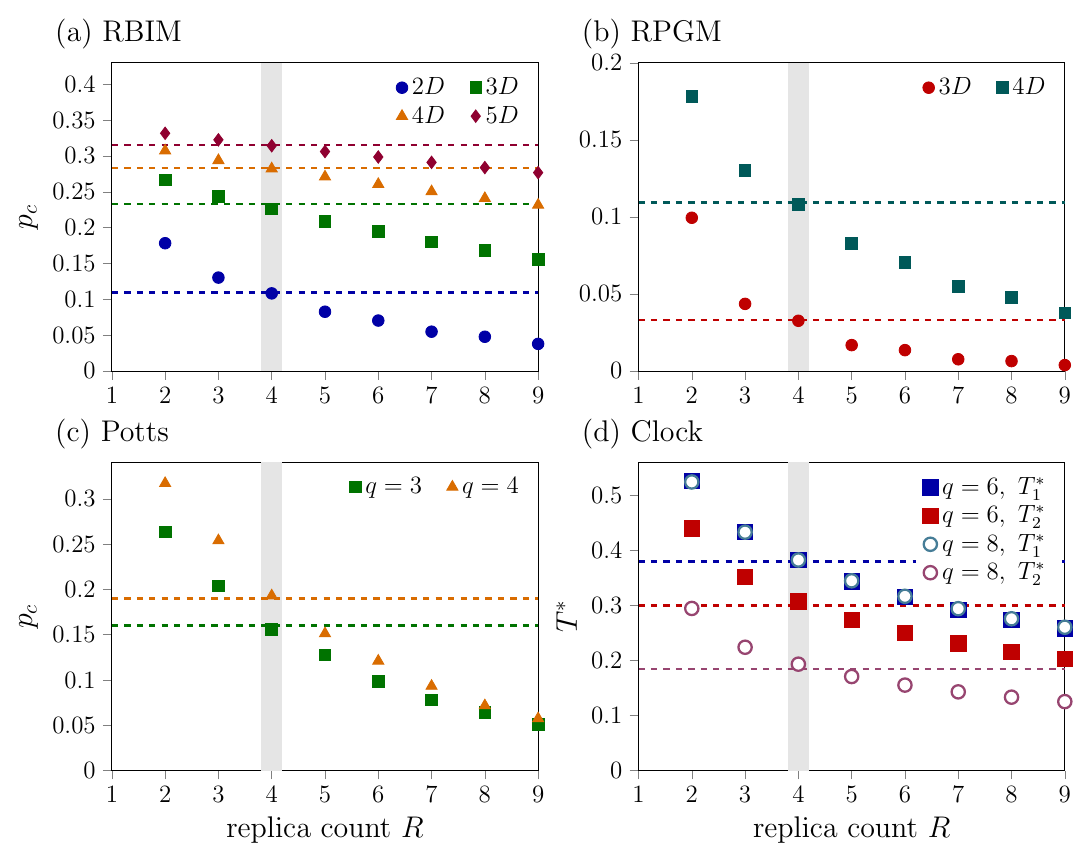}
	\caption{
		{\bf Threshold as a function of replica number $\mathbf{R}$.}
		For the variety of models of Table~\ref{tab:thresholds}, we plot the Nishimori threshold estimate
		as a function of Replica count that we use for our Fourier--Walsh projection.
		Consistently, we find that the $R=4$ estimate (highlighted by the gray bars) 
		matches the existing numerical values (dashed horizontal lines).
        }
	\label{fig:R_anlysis}
\end{figure}

This two, three, and four-replica structure is
standard, the propagators of replica field theory are linear combinations
of $\mathbb{E}[\langle\sigma\sigma\rangle^2]$,
$\mathbb{E}[\langle\sigma\sigma\rangle\langle\sigma\rangle\langle\sigma\rangle]$,
and $\mathbb{E}[\langle\sigma\rangle^2\langle\sigma\rangle^2]$, and
spin-glass simulations build unbiased overlap and vertex estimators from
several real replicas per disorder sample for closely related reasons
\cite{PhysRevB.88.224416,PhysRevE.109.055302}.
The local projection mirrors this overlap anatomy. Factoring the Ising
measurement sum gives
\begin{equation}
W_R =
2\cosh^R\!\eta
\sum_{s=0}^{\lfloor R/2\rfloor}
\gamma^{2s}\,X_{2s}\,,
\label{eq:WR}
\end{equation}
with $X_0=1$ and
$X_{2s}=\sum_{1\le a_1<\cdots<a_{2s}\le R} x_{a_1}\cdots x_{a_{2s}}$.
At $R=2$ only the pair channel $X_2$ is present. At $R=4$ the algebra
additionally contains the four-replica character
$x_1x_2x_3x_4=(x_1x_2)(x_3x_4)$, the local representative of a product of
two disjoint pair channels. Hence $R=4$ is the smallest local algebra
containing both the EA pair channel and a disjoint-pair channel needed to
form its connected fluctuation. For Potts and clock variables the same
statement is implemented by the corresponding finite-group pair characters
and first-harmonic pair channels \footnote{
We want to emphasize that these are real replicas in the estimator sense of
Refs.~\cite{PhysRevB.88.224416,PhysRevE.109.055302}, as they count the degree of the
integer-moment observable, not the replica number of an averaged free
energy. On the Nishimori line the record average of any fixed product of
thermal averages is performed exactly at integer replica number, as in
Eq.~\eqref{eq:WR}, and no analytic continuation is invoked.
}. 
For all the models at hand, summarized in Fig.~\ref{fig:R_anlysis}, 
we can see that $R=4$ is indeed the ``sweet spot" 
where the MRP projection intersects the ``true" numerical value.
Note that our line of argument above is a {\sl minimality} prescription, not a convergence theorem. 
Larger replica numbers, $R>4$, retain higher local moments, but at the single-bond level they 
cannot encode the lattice geometry to which those moments couple. In the examples
of Fig.~\ref{fig:R_anlysis}, they are instead  found empirically to  over-dress the pair channel.

Our four-replica construction also extends beyond hypercubic lattices (where all elementary loops are of length 4).
To illustrate this point we consider a hexagonal lattice geometry in Appendix~\ref{app:Honeycomb} 
and show that, again, the $R=4$ replica calculation produces a close estimate of the Nishimori transition.
The key observation to obtain a faithful estimate is that for the honeycomb lattice the elementary
unit cell has two sites and the {\sl single-bond} projection scheme should be adjusted to a {\sl cell} projection 
scheme. In analogy with the star--triangle map~\cite{baxter1985exactly} of the clean model one traces out one sublattice 
exactly in this cell projection approach to obtain an $R=4$ threshold estimate of $p_c\simeq 6.91\%$
in good approximation of the numerical~\cite{ZhuTantivasadakarnVishwanthTrebstVerresen} threshold estimate $p_c^{\rm num}\simeq 6.75\%$,
whereas a single-bond projection scheme would have resulted in an estimate that is some $30\%$ off, see Appendix~\ref{app:Honeycomb}.
 


\paragraph{Discussion.---} 
Our minimal-replica projection scheme to locate Nishimori thresholds 
builds on an  exact local decomposition \eqref{eq:treeplusloop} followed by a one-channel critical ansatz.
The crucial ingredient from Nishimori universality are the emergent 
local gauge identities \eqref{eq:Nishimori} that lock overlap fluctuations 
to the ordering channel, which motivates our ansatz to interpret
the Fourier--Walsh projected pair-channel coupling as the effective
coupling of this common pair or ``replicon"~\cite{Almeida_1978,Bray_1979} 
channel and locate criticality by matching it to the corresponding clean
critical coupling. 
In the replica theory of spin glasses \cite{mezard1988spin}, it is long 
established that the pair-overlap {\sl fluctuations} decompose, 
under the replica permutation symmetry  $S_R$, 
into longitudinal, anomalous, and replicon sectors of dimensions $1$,
$R-1$, and $R(R-3)/2$, respectively~\cite{Almeida_1978,Bray_1979,sk}. 
The de~Almeida--Thouless stability boundary of spin glasses
is then marked by the vanishing of the replicon eigenvalue
(of the replica-symmetric $R \times R$ Hessian) 
\footnote{At the replica-symmetric Sherrington-Kirkpatrick saddle~\cite{sk}, the overlap Hessian is specified
by three matrix elements, $P$, $Q$, and $R$, according to whether two
replica pairs share two, one, or no indices
\cite{Almeida_1978}.
The replicon eigenvalue is
$\lambda_{\mathrm{rep}}=P-2Q+R$, with degeneracy $n(n-3)/2$ (here, $R$ is the matrix element and $n$ is the replica count) and hence
two independent modes at $n=4$, $\lambda_{\mathrm{rep}}
 =1-\left(\frac{J}{kT}\right)^2
  (1-2q_{\mathrm{SK}}+r_{\mathrm{SK}}),$
where $q_{\mathrm{SK}}$ and $r_{\mathrm{SK}}$ are the two- and
four-replica moments. This structure motivates the minimal $R=4$
closure: four replicas are first required to form two disjoint pair
characters, whose quartic contribution dresses the pair coupling as
$K_{\mathrm{MRP}}=\kappa_2+K_4$.}.
Note that the pair-character subspace contains no replicon sector 
for replica count $R<4$, while at $R=4$ it has dimension two. These facts
motivate, in the context of conventional spin glass theory, to employ 
$R=4$ as the minimal auxiliary replica algebra.
Here we arrive at a similar conclusion, but note that along the Nishimori line 
considered in our context, the gauge identities enforce that 
overlap and ordering susceptibilities soften together rather than
through an independent de Almeida--Thouless instability. 

While for our minimal replica projection scheme we typically observe agreement on the percentage-point level,
the single largest
deviation is found for the 3D RBIM -- precisely the case in which the
disorder is {\sl Harris-relevant} at the clean critical point, so that the
local dressing is expected to be least faithful; the same argument applied to
the clock models (where there is no Harris-relevant non-Gaussian channels) is consistent with no significant deviation 
of our estimates to the existing numerical data (beyond the percentage-point precision). 
The same perspective also explains the observed failure mode.  The matching
to \(\beta_c^{\rm clean}\) assumes a continuous clean critical channel.
For 2D Potts models this holds for \(q\le4\), while for
\(q>4\) the clean transition turns first order and
\(\beta_c^{\rm clean}\) is a coexistence point rather than a
fixed-point coupling.  The resulting drift of the naive \(q>4\)
estimates is therefore a structural breakdown of the one-channel
projection, not a separate anomaly.

Beyond the Ising/Potts hierarchy, the clock model provides the sharpest test
of the projection.  Because the clean \(\mathbb{Z}_q\) clock model has
two BKT transitions for \(q\ge5\), the projection inherits two clean
critical inputs and produces two Nishimori temperatures.  This is
unlike a single self-dual scalar condition, which can select at most
one point.  The agreement with recent \(\mathbb{Z}_q\) toric-code
numerics~\cite{VijayLee2025}, together with the variance decomposition
showing \(94\)--\(96\%\) weight in the first harmonic, indicates that
the retained first harmonic captures the dominant local Nishimori
channel.  
The approximate satisfaction of the Gilbert--Varshamov
self-dual entropy relation provides an independent check; 
notably,  in the large-$q$ limit the ratio $\left(H_q(T_1^\ast) + H_q(T_2^\ast)\right)/\ln q$ approaches unity
(as derived in  Appendix \ref{app:ZqModels}) 
implying the restoration of an {\it approximate} self-duality in the large-$q$ limit. 

This leaves us with the tantalizing observation that while one typically does not expect to make any connections between 
a pair of critical theories for a clean and disordered model (described by the $R\to 0$ replica limit) beyond universality classes 
(and related RG flows), the situation for a pair of clean and {\sl Born-disordered} model (described by the $R\to 1$ replica limit)
might allow for a deeper intrinsic connection that also allows one to make quantitative statements about the location of 
critical points/thresholds and dualities. \\

\paragraph{Data availability.---} 
The data provided in the various Tables and Plots  are available on Zenodo~\cite{zenodo_MRP}.\\

\paragraph{Acknowledgments.---} 

We are indebted to A.~Ludwig, R.~Patil, and G.-Y.~Zhu for stimulating this work and collaboration on several closely related projects. R.M.\ acknowledges support from a postdoctoral fellowship from the Alexander von Humboldt Foundation.
We also thank R. Andrist for sharing their numerical data.
The Cologne research group is supported, in part, by
the Deutsche Forschungsgemeinschaft (DFG, German Research Foundation) under Germany’s Excellence Strategy—Cluster of Excellence Matter
and Light for Quantum Computing (ML4Q) EXC 2004/1 -- 390534769 
as well as within the CRC network TR 183 (Project Grant No.\ 277101999) as part of subproject B01.
Our numerical simulations were performed on the Otus cluster at PC2 in Paderborn and the RAMSES cluster at RRZK Cologne. 
%

\appendix

\section{Potts models}
\label{app:PottsModels}

\subsection*{Replicated bond weights}
For $q$-state Potts variables with $R=4$ replicas, the local
replicated bond log-weight depends on the occupation vector
$N=(N_0,\ldots,N_{q-1})$ with $\sum_n N_n = 4$, where $N_n$
counts the number of replicas with bond color $n$.  Each sector contributes \(W_N\) configurations to the uniform group
average. The order-parameter channel is the pair-coincidence count
\[
    A_2 = \sum_n \binom{N_n}{2} \,,
\]
which counts the number of replica pairs occupying the same Potts bond
state, and has exact log-weight
$y_N = \ln\sum_{n=0}^{q-1} e^{J_0 N_n}$ (see Table~\ref{tab:Potts-sectors}).
Projecting onto $A_2$ via Fourier--Walsh gives for $q=3$

\[
        K_{\rm MRP}^{q=3  \rm Potts}
        =
        \frac{3y_A+6y_B-9y_D}{27}
        =
        \frac{y_A+2y_B-3y_D}{9} \,,
\]
and for $q=4$,
\begin{equation*}
	\label{eq:potts4}
	K_{\rm MRP}^{q=4 \; \rm Potts} = (y_A+4y_B+y_C-4y_D-2y_E)/16 \,.
\end{equation*}

The Nishimori threshold is the measurement strength at which the projected
coupling reaches the clean Potts critical point,
\begin{equation}
  K_{\rm MRP}^{q\text{-}\mathrm{Potts}}(J_0^{*})=\beta_c(q)=\ln\!\big(1+\sqrt q\big),
  \label{eq:Potts-match}
\end{equation}
solved for $J_0^{*}$ and converted through
\begin{equation}
  \gamma_c=\frac{e^{J_0^{*}}-1}{e^{J_0^{*}}+q-1},
  \qquad
  p_c=\frac{q-1}{q}\,(1-\gamma_c).
  \label{eq:Potts-convert}
\end{equation}
This gives $J_0^{*}=2.3837$, $p_c\simeq15.6\%$ for $q=3$ and
$J_0^{*}=2.5294$, $p_c\simeq19.3\%$ for $q=4$, the values quoted in
Table~\ref{tab:thresholds}.

\begin{table}[t]
	\caption{{\bf Potts sectors} for the $q$-state Potts model at $R=4$.}
	\label{tab:Potts-sectors}
	\centering
	\begin{ruledtabular}
	\begin{tabular}{ccccc}
	& $N$ & $W_N$ & $x_N$ & $y_N$ \\  
	\hline
	$q=3$ 	& $A=(4,0,0)$ & 3 & 6 & $\ln(e^{4J_0}+2)$ \\ 
			& $B=(3,1,0)$ & 24 & 3 & $\ln(e^{3J_0}+e^{J_0}+1)$ \\
			& $C=(2,2,0)$ & 18 & 2 & $\ln(2e^{2J_0}+1)$ \\ 
			& $D=(2,1,1)$ & 36 & 1 & $\ln(e^{2J_0}+2e^{J_0})$ \\ 
	\hline
	$q=4$	& $A=(4,0,0,0)$ & 4 & 6 & $\ln(e^{4J_0}+3)$\\
			& $B=(3,1,0,0)$ & 48 & 3 & $\ln(e^{3J_0}+e^{J_0}+2)$\\
			& $C=(2,2,0,0)$ & 36 & 2 & $\ln(2e^{2J_0}+2)$\\
			& $D=(2,1,1,0)$ & 144 & 1 & $\ln(e^{2J_0}+2e^{J_0}+1)$\\
			& $E=(1,1,1,1)$ & 24 & 0 & $\ln(4e^{J_0})$\\
	\end{tabular}
	\end{ruledtabular}
\end{table}

\subsection*{Nature of transition and break-down of projection scheme}

\begin{figure}[t]
    \centering
    \includegraphics[width=\columnwidth]{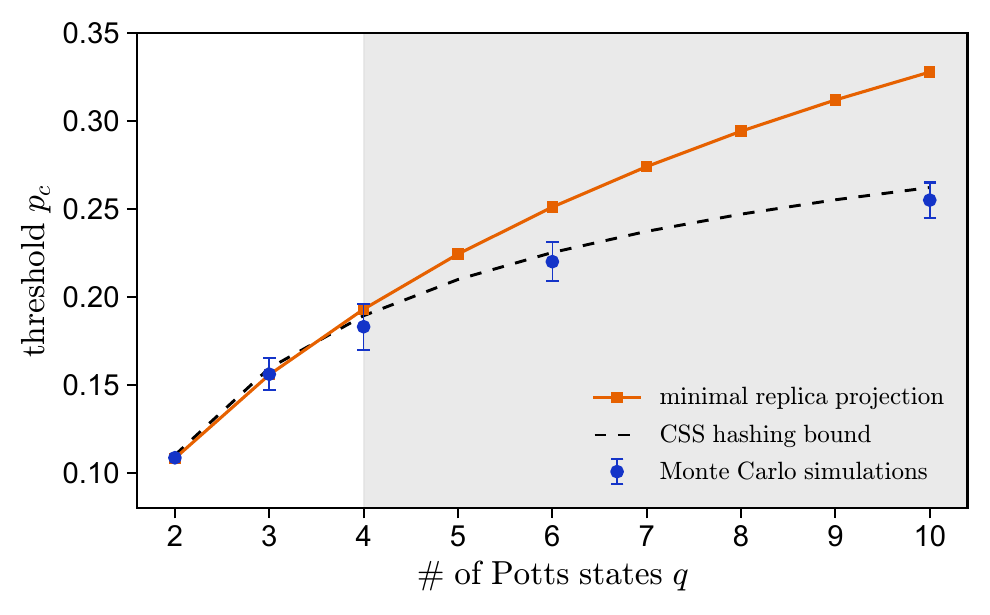}
    \caption{{\bf Nishimori thresholds for $\mathbf{q}$-state Potts models.}
    		Comparison of error correction thresholds obtained by the Hashing bound,
		our MRP estimations and numerical results~\cite{Poulin2013,Andrist2015}. 
		The shaded region $q\ge 5$ indicates the regime where the clean
		Potts transition is first-order, outside the validity of the projection construction.}
\label{fig:Potts-Hashing}
\end{figure}

For $q \le 4$ the clean 2D Potts transition is
continuous ($q=4$ marginal), and matching the projected
coupling to $\beta_c^{\rm clean}$ is anchored to a genuine
critical fixed point.  For $q>4$ the clean Potts transition
is first-order and $\beta_c^{\rm clean}$ marks a coexistence
point of two thermodynamic branches rather than a continuous
critical scaling field, so the single scalar projection has
no fixed-point coupling to inherit from the clean theory.
This does not mean the disordered model has no continuous transition:
quenched randomness  can round first-order behavior in 2D 
and produce a continuous random fixed
point~\cite{aizenman1990rounding,PhysRevLett.79.4063,duminil2017continuity}.  The
transition is then controlled by disorder-generated physics
that the one-coordinate projection cannot capture.
Consequently the $q>4$ MRP estimates should be regarded as
uncontrolled; their drift above both the hashing bound and
the available numerical thresholds, compared  in Fig.~\ref{fig:Potts-Hashing},
is a diagnostic of this breakdown.  
We have marked this regime by the shaded region in the figure.

This limitation is analogous to what happens in other
single-coupling approaches. The conjectured principal Boltzmann-factor
duality by Nishimori and Nemoto~\cite{NishimoriDualityConjecture} compresses the replicated
weight to a single scalar self-dual condition.
Similarly,
a Bethe/cavity linearization locates the point where the
paramagnetic fixed point becomes locally unstable.  For a
continuous transition this instability coincides with the
thermodynamic transition, but for a first-order transition (Potts on the Bethe lattice 
at $q\ge 3$~\cite{mezard2009information}) it
only gives a spinodal; the actual transition must instead be
found by comparing the free energies of the competing
branches.  In all cases the limitation is the same -- a
single-coupling matching is meaningful for a continuous fixed
point, but not for a first-order clean transition where the
thermodynamic transition is determined by branch competition.

\subsection*{Comparison to CSS hashing bound}
\begin{figure}[t]
    \centering
    \includegraphics[width=1.0\columnwidth]{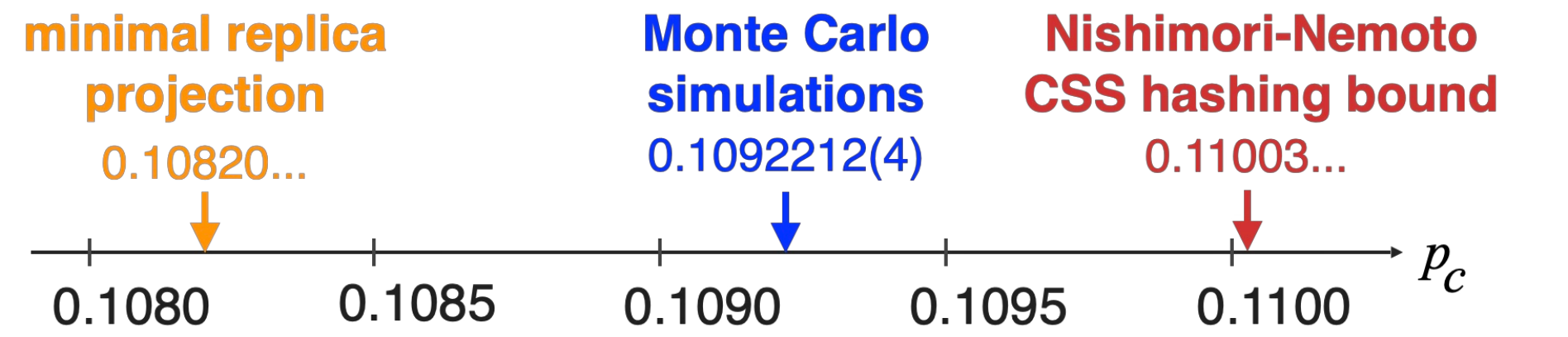}
  \caption{{\bf Threshold comparison} for the 2D surface code.
	Shown are the estimates obtained from our projection method, 
	large-scale Monte Carlo simulations~\cite{b8y5-k3y6}, 
	coinciding CSS-hashing~\cite{PhysRevA.54.1098} and Nishimori--Nemoto~\cite{NishimoriDualityConjecture} values. 
	The latter two fall onto the same point because both conditions reduce to $H_2(p)=\tfrac12$ at $q=2$.
	The error bar of the numerical estimate is smaller than the line width.
		}
    \label{fig:ThresholdComparison}
\end{figure}

It is useful to compare our threshold estimates to the CSS hashing
benchmark~\cite{PhysRevA.54.1098,BennettDiVincenzoSmolinWootters1996}. For a
$q$-ary symmetric error channel, let
\[
	H_q(p)=-(1-p)\log(1-p)-p\log\frac{p}{q-1}
\]
be the error entropy.  The zero-rate CSS hashing condition is $H_q(p_{\rm hash})=\frac12\log q$
(which, for $q=2$ is identical to the conjectured Nishimori-Nemoto duality). 
For binary errors this gives $p_{\rm hash}(2)\sim 0.11003...$ 
-- a value that is similarly close to the numerically obtained (true) threshold as our MRP estimate,
but while the latter {\sl underestimates} the true value, the hashing bound {\sl overestimates} it,
see the comparison of Fig.~\ref{fig:ThresholdComparison}.
The $\mathbb{Z}_q$ toric code
under the symmetric bit-flip channel -- the natural qudit
generalization of the binary symmetric channel -- maps onto the
$q$-state random-bond Potts model on the Nishimori
line~\cite{Andrist2015,Poulin2013}, so the hashing condition
above and our Potts thresholds in Table~\ref{tab:thresholds} describe the same
class of decoding problems.  Strictly speaking, topological
codes are degenerate, so the hashing value is {\it not} a
rigorous upper bound on the error threshold; 
similarly we cannot rigorously show that our projection estimate is a lower bound on the error threshold.

\section{$\mathbb{Z}_q$ clock models}
\label{app:ZqModels}

For the \(q\)-state clock model define the {\it von-Mises distribution}~\cite{mardia2009directional} at Nishimori temperature \(T=1/J\) ($J$ controls the measurement strength in the learning protocol),
\[
        p_k(T)
        =
        \frac{\exp\!\left[T^{-1}\cos(2\pi k/q)\right]}
        {\sum_{\ell=0}^{q-1}\exp\!\left[T^{-1}\cos(2\pi \ell/q)\right]},
        \qquad
        k=0,\ldots,q-1,
\]
and the Shannon entropy,
\[
        H_q(T)=-\sum_{k=0}^{q-1}p_k(T)\ln p_k(T).
\]
For \(q\le4\), where there is a single transition, so the \textit{Hashing} type bound for the $q$-state clock model becomes,
\[
        \ln q \simeq 2H_q(T^\ast).
\]
For \(q\ge5\), with two BKT thresholds, the conjectured~\cite{VijayLee2025} self-duality relation is,
\[
        \ln q \simeq H_q(T_1^\ast)+H_q(T_2^\ast).
\]

The finite-\(q\) clock projection has a simple continuum limit.  The clock angles become continuous, and the discrete average over the local measurement angle becomes an angular integral.  Fix one replica angle to zero by translation invariance and define
\[
        \rho(\theta)
        =
        \left|1+\ee^{\ii\theta_1}+\ee^{\ii\theta_2}
        +\ee^{\ii\theta_3}\right|,
        \qquad
        X(\theta)=\frac{\rho(\theta)^2-4}{2}.
\]
The discrete disorder sum over the clock noise angle becomes a Bessel function,
\[
        \frac1q\sum_{m=0}^{q-1}
        \exp\left[J\Ree\!\left(\ee^{2\pi\ii m/q}
        \sum_a\ee^{\ii\theta_a}\right)\right]
        \longrightarrow
        I_0\!\left(J\rho(\theta)\right),
\]
so the large-\(q\) first-harmonic projected coupling is
\begin{equation}
K_\infty(J)
        =
        \frac{1}{3(2\pi)^3}
        \int_0^{2\pi}\!\!\int_0^{2\pi}\!\!\int_0^{2\pi}
        X(\theta)\ln I_0\!\left(J\rho(\theta)\right)
        \,d\theta_1\,d\theta_2\,d\theta_3 .
\label{eq:kinfty}
\end{equation}

From this, large-\(q\) lower Nishimori clock threshold can be calculated as,
\begin{equation}
J_2^\ast(q)
        \sim
        \frac{q^2}{8\pi c_\infty}
        =
        \frac{q^2}{12.249325},
        \qquad
        T_2^\ast(q)
        \sim
        \frac{12.249325}{q^2}.
\label{eq:largeqlower}
\end{equation}

Insert the large-\(q\) clock Nishimori threshold points into the clock entropy condition
\[
        H_q(T_{\mathrm{hi}}^\ast)+H_q(T_2^\ast(q))
        \stackrel{?}{\simeq}
        \ln q.
\]

It can be shown approximately,

\begin{equation}
H_q(T_{\mathrm{hi}}^\ast)+H_q(T_2^\ast(q))
        =
        \ln q +\mathcal{O}(1).
\end{equation}

So, large $q$ limit (where the ratio approaches 1), we have shown that our projection method recovers the GV entropy duality condition.

\section{Honeycomb lattice}
\label{app:Honeycomb}

The single-bond criterion of the main text keeps only the information of one bond and
the scalar $\beta_c^{\mathrm{clean}}$; all other lattice structure is discarded
by the projection. This truncation gives few-percent estimates for the
hypercubic lattice thresholds of Table~I, but there is no bound on its error, and for the honeycomb
lattice it is off by $\sim30\%$ (below). Because the honeycomb is bipartite with
coordination three, one sublattice can be traced out exactly, mapping it
to the triangular lattice. We observe that projecting this exact cell, which retains the
honeycomb connectivity discarded at single-bond level, instead of a single
bond, removes most of the error.

\begin{figure}[htp]
	\centering
	\includegraphics[width=0.85\linewidth]{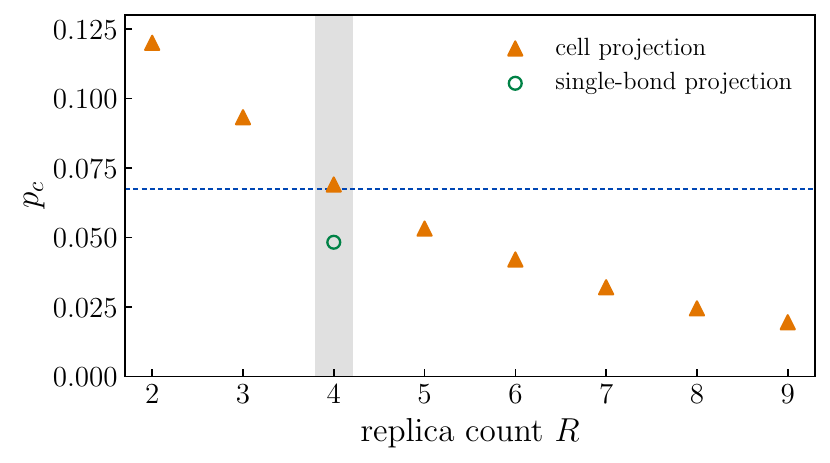}
	\caption{
		{\bf Threshold for hexagonal lattice geometry.}
		 We plot the error correction threshold across for honeycomb lattice as a function of replica count $R$ that we use for our Fourier--Walsh projection.
		 }
	\label{fig:R_anlysishoney}
\end{figure}

For the honeycomb lattice one sublattice is therefore traced out before
projecting. In the clean input model this is the well known star-triangle transformation~\cite{baxter1985exactly},
which maps the honeycomb critical point
$\beta_c^{\mathrm{clean}}=\tfrac12\ln(2+\sqrt3)$ to the triangular critical
point $\beta_c^{\mathrm{clean}}=\tfrac14\ln3$. So the decimated cell must be
matched to the triangular $\beta_c^{\mathrm{clean}}=\tfrac14\ln3$.

On the Nishimori line ($\beta=0$) the replicated weight of each bond is
$2\cosh(\eta\,\sigma_0\!\cdot\!\sigma_i)$ with
$\sigma_0\!\cdot\!\sigma_i=\sum_{a=1}^{R}\sigma_0^{(a)}\sigma_i^{(a)}$ (main
text), $R=4$. For the spin $\sigma_0$ shared by the three bonds
$(\sigma_1,\sigma_2,\sigma_3)$, tracing out $\sigma_0$ gives the cell
log-weight
\begin{equation}
  y(\sigma_1,\sigma_2,\sigma_3)
  =\ln\!\sum_{\{\sigma_0^{(a)}=\pm1\}}\;
   \prod_{i=1}^{3}2\cosh(\eta\,\sigma_0\!\cdot\!\sigma_i).
\end{equation}
The cell inner product is the uniform average over the three external spins,
and the pair-overlap channel of the tested pair $(\sigma_1,\sigma_2)$ is
\begin{equation}
  \langle f,g\rangle
  =2^{-3R}\!\!\sum_{\sigma_1,\sigma_2,\sigma_3\in\{\pm1\}^{R}}\!\! f\,g,\ X_2=\sum_{a<b}\sigma_1^{(a)}\sigma_2^{(a)}\sigma_1^{(b)}\sigma_2^{(b)} .
\end{equation}
(By the symmetry of the three legs the choice of tested pair is immaterial.)
Projecting $y$ onto $X_2$ and matching to the triangular
$\beta_c^{\mathrm{clean}}$,
\begin{equation}
  \frac{\langle y,X_2\rangle}{\langle X_2,X_2\rangle}=\tfrac14\ln3 ,
\end{equation}
gives
\begin{equation}
  \eta_c=1.30028,\ \gamma_c=\tanh\eta_c=0.86180,\ p_c=\tfrac{1-\gamma_c}{2}=6.91\% .
\end{equation}
The single-bond projection matched directly to the honeycomb
$\beta_c^{\mathrm{clean}}=\tfrac12\ln(2+\sqrt3)$ gives instead $4.83\%$, low by
$\sim30\%$ relative to the numerical threshold $\simeq6.75\%$~\cite{ZhuTantivasadakarnVishwanthTrebstVerresen}. 
The trace does not close for a spin shared by four bonds, eliminating it
generates a four-spin interaction $\sigma_1\sigma_2\sigma_3\sigma_4$ (an even
character, so it does not vanish), and the result is not a pair lattice. No
exact star-triangle map exists in that case, so we do not apply this cell
projection there.

\bibliography{Thresholds}

\clearpage


\makeatletter
\let\@sectioncntformat\@seccntformat
\def\@hangfrom@section#1#2#3{\@hangfrom{#1#2}\MakeTextUppercase{#3}}
\def\@hangfroms@section#1#2{#1\MakeTextUppercase{#2}}
\makeatother

\renewcommand{\thesection}{S\arabic{section}}
\renewcommand{\theequation}{S\arabic{equation}}
\renewcommand{\thefigure}{S\arabic{figure}}
\renewcommand{\thetable}{S\arabic{table}}

\begin{widetext}
\begin{center}
  \textbf{\large Supplemental Material:\\[2pt]
  Nishimori Threshold Estimation for Bayesian Inference
  and $\mathbb{Z}_q$ Surface Code Decoding}\\[8pt]
  Rohit Mukherjee and Simon Trebst\\
  \textit{Institute for Theoretical Physics, University of Cologne,
  Z\"ulpicher Stra\ss e 77, 50937 Cologne, Germany}
\end{center}
\end{widetext}

\clearpage

\setcounter{table}{0}
\renewcommand{\thetable}{S\arabic{table}}

\renewcommand{\thesection}{S\arabic{section}}
\setcounter{section}{0}

\begin{table*}[h]
\caption{\textbf{Clean critical couplings $\beta_c^{\rm clean}$} used as inputs for each row of Table~\ref{tab:thresholds} in the main text.
Values marked {\it exact} follow from Onsager's solution or
Kramers--Wannier--Wegner / Potts self-duality. {\it Exact dual} entries are
inherited from the dual lattice's value. {\it MC} entries are best-published
Monte Carlo determinations. Clock $\beta_{c1}(q\!\ge\!5)$ saturates to the XY
BKT value; $\beta_{c2}(q)$ follows the JKKN asymptotic $q^2/(8\pi)$ measured
numerically at finite $q$. }
\label{tab:clean_inputs}
\begin{ruledtabular}
\begin{tabular}{l l l c}
Model / setting & $\beta_c^{\rm clean}$ & approach & Ref. \\
\hline
Ising 2D                                 					& $\tfrac{1}{2}\ln(1\!+\!\sqrt{2}) = 0.4406868...$ & exact (Onsager)         & \cite{PhysRev.65.117} \\
Ising 3D                           				& $0.22165462(2)$                                & MC                      & \cite{Ferrenberg2018} \\
Ising 3D PGM     			& $0.76141331(4)$                                   & exact dual of 3D Ising  & \cite{Wegner1971} \\
Ising 4D                           					& $0.14969378(15)$                                   & MC                      & \cite{LUNDOW2023116256} \\
Ising 4D PGM              			& $0.4406868...$\,{\small(= 2D Ising)}              & exact dual of 2D RBIM   & \cite{PhysRev.65.117} \\
Ising 5D                           					& $0.1139150(4)$                                 & MC                      & \cite{LUNDOW2015305} \\
\hline
Potts 2D $q\!=\!3$              				& $\ln(1\!+\!\sqrt{3}) = 1.00505...$                & exact   & \cite{Wu1982} \\
Potts 2D $q\!=\!4$              				& $\ln 3 = 1.09861...$                              & exact  & \cite{Wu1982} \\
\hline
Clock $q\!=\!2$ \,($\equiv$ Ising)       				& $\tfrac{1}{2}\ln(1\!+\!\sqrt{2}) = 0.4406868...$ & exact (Onsager)         & \cite{PhysRev.65.117} \\
Clock $q\!=\!4$ \,($\equiv 2\times$Ising) 			& $\ln(1\!+\!\sqrt{2}) = 0.8813735...$              & exact & \cite{Wu1982,Jose1977} \\

Clock $q\!=\!5$, upper BKT $\beta_{c1}$  			& $1.05031(22)$       & MC       & \cite{PhysRevE.101.060105} \\
\phantom{Clock $q\!=\!5$, }lower BKT $\beta_{c2}$  	& $1.10387(24)$       & MC       & \cite{PhysRevE.101.060105} \\

Clock $q\!=\!6$, upper BKT $\beta_{c1}$  			& $\beta_{\rm BKT}^{\rm XY} \simeq 1.09565(60)$       & MC (XY BKT limit)       & \cite{PhysRevE.101.060105,Jose1977} \\
\phantom{Clock $q\!=\!6$, }lower BKT $\beta_{c2}$  	& $1.44907(84)$ \,{\small (JKKN: $36/8\pi\!=\!1.43$)}    & MC + JKKN               & \cite{PhysRevE.101.060105,Jose1977} \\

Clock $q\!=\!7$, upper BKT $\beta_{c1}$  			& $1.10241(61)$       & MC       & \cite{PhysRevE.101.060105} \\
\phantom{Clock $q\!=\!7$, }lower BKT $\beta_{c2}$  	& $1.88501(107)$      & MC       & \cite{PhysRevE.101.060105} \\

Clock $q\!=\!8$, upper BKT $\beta_{c1}$ 			 & $\beta_{\rm BKT}^{\rm XY} \simeq 1.10375(61)$       & MC (XY BKT limit)       & \cite{PhysRevE.101.060105,Jose1977} \\
\phantom{Clock $q\!=\!8$,} lower BKT $\beta_{c2}$  & $2.39693(172)$ \,{\small (JKKN: $64/8\pi\!=\!2.55$)}    & MC + JKKN               & \cite{PhysRevE.101.060105,Jose1977} \\
\hline
XY 3D                            				& $0.4541652(11)$  & MC        & \cite{PhysRevB.100.224517} \\
XY 4D                            				& $0.30171037(55)$  & MC        & \cite{lv2021finite} \\

\end{tabular}
\end{ruledtabular}
\end{table*}

\begin{table*}[t]
\caption{\textbf{Projected Nishimori thresholds} computed from the clean
\(\beta_c^{\rm clean}\) inputs of Table~\ref{tab:clean_inputs}. For each model, we list the Nishimori inverse temperature $\beta_N$ 
(or coupling $J_0^*$), the corresponding critical measurement strength 
$\gamma_c$, and the location of the transition, expressed either as a 
critical error rate $p_c$ (Ising / Potts models) or as a critical 
temperature $T^*$ ($T_{1,2}^*$ for the upper and lower BKT transitions 
of the clock models).}
\label{tab:S2_projected_thresholds}
\centering
\begin{ruledtabular}
\begin{tabular}{@{}llllr@{}}
Model / setting &
\multicolumn{1}{c}{\(\beta_N\) or $J_{0}^{*}$} &
\multicolumn{1}{c}{\(\gamma_c\)} &
\multicolumn{1}{c}{transition}\\
\hline
Ising 2D
& 1.054606205...
& 0.783590714...
& \(p_c=10.8205...\%\)\\
Ising 3D
& \(0.61477461(4)\)
& \(0.54747934(3)\)
& \(p_c=22.6260332(14)\%\)\\
Ising 3D PGM
& \(1.69611309(9)\)
& \(0.93492140(1)\)
& \(p_c=3.2539301(6)\%\)\\
Ising 4D
& \(0.46677135(31)\)
& \(0.43558694(25)\)
& \(p_c=28.220653(13)\%\)\\
Ising 4D PGM
& 1.054606205...
& 0.783590714...
& \(p_c=10.8205...\%\)\\
Ising 5D
& \(0.3903433(9)\)
& \(0.3716561(8)\)
& \(p_c=31.417193(38)\%\)\\
\hline
Potts \(q=3\)
& 2.383653584...
& 0.766436118...
& \(p_c=15.5709...\%\)\\
Potts \(q=4\)
& 2.529369998...
& 0.742692464...
& \(p_c=19.2981...\%\)\\
\hline
Clock \(q=2\) (\(\equiv\) Ising)
& 1.054606205...
& 0.783591...
& \(T^\ast=0.948221...\)\\
Clock \(q=4\) 
& 2.109212410...
& 0.783591...
& \(T^\ast=0.474111...\)\\
Clock \(q=5\), upper BKT \(\beta_{c1}\)
& \(2.49938(46)\)
& 0.793033(61)
& \(T_1^\ast=0.400099(73)\)\\
\phantom{Clock \(q=5\),} lower BKT \(\beta_{c2}\)
& \(2.61104(50)\)
& $\quad$ 0.807252(61)
& \(T_2^\ast=0.382990(74)\)\\
Clock \(q=6\), upper BKT \(\beta_{c1}\)
& \(2.58334(123)\)
& 0.781768(136)
& \(T_1^\ast=0.387096(184)\)\\
\phantom{Clock \(q=6\),} lower BKT \(\beta_{c2}\)
& \(3.30998(173)\)
& $\quad$ 0.846369(123)
& \(T_2^\ast=0.302117(158)\)\\
Clock \(q=7\), upper BKT \(\beta_{c1}\)
& \(2.59545(125)\)
& 0.776771(133)
& \(T_1^\ast=0.385290(185)\)\\
\phantom{Clock \(q=7\),} lower BKT \(\beta_{c2}\)
& \(4.20287(221)\)
& $\quad$ 0.880307(90)
& \(T_2^\ast=0.237933(125)\)\\
Clock \(q=8\), upper BKT \(\beta_{c1}\)
& \(2.59700(124)\)
& 0.775482(133)
& \(T_1^\ast=0.385060(184)\)\\
\phantom{Clock \(q=8\),} lower BKT \(\beta_{c2}\)
& \(5.24423(354)\)
& $\quad$ 0.904853(87)
& \(T_2^\ast=0.190686(129)\)\\
\hline
XY 3D
& \(1.2706676(23)\)
& 0.5341255(7)
& \(T^\ast=0.7869879(14)\)\\
XY 4D
& \(0.9510223(12)\)
& 0.4287662(4)
& \(T^\ast=1.0515000(13)\)\\
\end{tabular}
\end{ruledtabular}
\end{table*}


In this Supplemental Material, we provide additional details for the
calculations presented in the main text. In Sec.~\ref{Supp:Ising}, we
review the Fourier--Walsh (FW) transform and derive the Ising pair-overlap
projection in full. In Sec.~\ref{Supp:PottsModels}, we describe the Potts
Fourier--Walsh characters and the corresponding pair-overlap channel. In
Sec.~\ref{Supp:ZqClockModels}, we derive the FW projection for the
$Z_q$ clock model, including its large-$q$ limit. In
Sec.~\ref{Supp:Uncertainty}, we explain how uncertainties in the projected
Nishimori thresholds are propagated. 
Next, in Sec.~\ref{sec:varianceprojection} we do the variance decomposition and show by single channel ansatz how much weight we are discarding.
Finally, in
Sec.~\ref{Supp:Bethe}, we discuss the Bethe-lattice calculation for Ising and Potts modesl.

\section{Ising Models}
\label{Supp:Ising}

The Fourier--Walsh expansion is the Fourier transform for functions
on the hypercube.
Let
\[
        x=(x_1,\ldots,x_R)\in\{\pm1\}^R .
\]
The natural inner product of two functions \(f,g:\{\pm1\}^R\to\mathbb R\) is the uniform average
\[
        \langle f,g\rangle
        =
        2^{-R}\sum_{x_1,\ldots,x_R=\pm1}f(x)g(x).
\]
For every subset \(A\subseteq\{1,\ldots,R\}\), define the Walsh character
\[
        \chi_A(x)=\prod_{a\in A}x_a,
        \qquad
        \chi_{\varnothing}(x)=1 .
\]
These characters are orthonormal:
\[
        \langle \chi_A,\chi_B\rangle=\delta_{A,B}.
\]
Therefore any function \(f(x)\) has the exact Walsh expansion
\begin{equation}
\begin{aligned}
    f(x) &= \sum_{A\subseteq\{1,\ldots,R\}} \widehat{f}_A\,\chi_A(x),\\
    \widehat{f}_A &= \langle f, \chi_A\rangle = 2^{-R}\sum_x f(x)\prod_{a\in A} x_a.
\end{aligned}
\end{equation}
The coefficient \(\widehat f_A\) measures the component of \(f\) in the \(A\)-replica channel.  
Thus\\
\[
\begin{array}{c|c}
|A|&\text{meaning}\\
\hline
0&\text{constant/normalization}\\
1&\text{one-replica field}\\
2&\text{pair-overlap channel}\\
4&\text{disjoint-pair channel, for }R\ge4 \,.
\end{array}
\]
For an Ising bond problem, the clean  channel is a two-replica pair, $x_ax_b$. So the effective clean coupling is the degree-two Walsh coefficient of the exact replicated log-weight. We now discuss the Ising bond energy measurement/ inference setting.

\subsection{Ising bond energy measurement/inference problem}
We now derive the replicated bond Hamiltonian for the Ising measurement problem at $\beta=0$ before applying the Fourier--Walsh projection. 
The Ising Hamiltonian has the form $H=-\sum_{ij} \sigma_{i}\sigma_{j}$.
Let the measured bond variable be a noisy binary record $m_{ij}=\pm1$ of the true Ising bond variable $\tau_{ij}=\sigma_i\sigma_j$. The measurement model/likelihood is~\cite{Patil2026,NahumJacobsen2025,PutzGarrattNishimoriTrebstZhu},
\begin{equation}
\label{eq:binary_likelihood_endmatter}
  P(m_{ij}|\sigma_i,\sigma_j)=\frac{e^{\eta m_{ij}\sigma_i\sigma_j}}{2\cosh\eta}
  =\frac{1+\gamma m_{ij}\sigma_i\sigma_j}{2},
  \ \gamma=\tanh\eta .
\end{equation}
At $\beta=0$ there is no prior thermal coupling between the spins, so the only local interaction in the posterior comes from the measurement likelihood.  For $R$ posterior replicas sharing the same measurement record, the replicated local likelihood is
\begin{equation}
\label{eq:replicated_likelihood_before_sum_endmatter}
  \prod_{a=1}^{R}P(m_{ij}|\sigma_i^{(a)},\sigma_j^{(a)})
  \propto
  \exp\left[\eta m_{ij}\sum_{a=1}^{R}\sigma_i^{(a)}\sigma_j^{(a)}\right].
\end{equation}

This is also proportional to the posterior distribution from \textit{Bayes' theorem} along the Nishimori line $\beta=0$.
The measurement outcome is not kept fixed in the averaged replicated theory; it is summed over as the shared measurement outcome.  Hence the exact single-bond measurement averaged replicated weight is
\begin{align}
\label{eq:ising_measurement_bond_weight_endmatter}
  W_R^{\rm Ising}(\{x_a\})
  &=\sum_{m_{ij}=\pm1}\exp\left(\eta m_{ij}\sum_{a=1}^{R}x_a\right)\nonumber\\
  &=2\cosh(\eta S),\ x_a=\sigma_i^{(a)}\sigma_j^{(a)},\ S=\sum_{a=1}^{R}x_a .
\end{align}
The additive constant $\ln2$ is irrelevant for the projected coupling, so the replicated bond log-weight/ effective Hamiltonian is
\begin{equation}
\label{eq:ising_measurement_log_weight_endmatter}
  y(S)=\ln\cosh(\eta S).
\end{equation}
This is the local object being projected. 
So, in general, before doing any Fourier--Walsh projection, we need the exact local replicated log-weight \(y\). The common structure is simple.  There is a local noisy measurement value \(m\).  For one replica with bond variable \(\Delta_a\), the likelihood has the form
\[
        P(m|\Delta_a)\propto \exp[-E(m,\Delta_a)] \,.
\]
For \(R\) replicas, the measurement/noise variable is shared, so the replicated local weight is
\[
        W_R(\Delta_1,\ldots,\Delta_R)
        =
        \sum_m \prod_{a=1}^R P(m|\Delta_a) \,.
\]
The object projected in the Fourier/Walsh calculation is
\[
        y_R(\Delta)=\ln W_R(\Delta),
\]
up to additive constants independent of \(\Delta\). In this notation, 

\begin{equation}
        y_R^{\rm Ising}(x)
        =
        \ln\cosh\left(\eta\sum_{a=1}^R x_a\right).
\label{eq:replicatedisingweight}
\end{equation}

\paragraph{Ising at \(R=4\): ordinary Walsh transform}
The Ising case is the cleanest place to see the whole mechanism.  The measurement variable couples to each replica through an Ising sign, and after summing over that measurement variable, the replicated bond weight depends only on the total replicated magnetization \(S=\sum_a x_a\). Since the pair Walsh function \(x_1x_2\) has zero uniform average, additive constants in the log-weight never affect the projected coupling.
For Ising variable, we write
\[
        x_a=\pm1,\qquad S=\sum_{a=1}^4x_a.
\]
The exact replicated measurement log-weight, up to an additive constant, is
\[
        y(x_1,x_2,x_3,x_4)=\ln\cosh(\eta S).
\]
The Walsh coefficient of the pair character \(x_1x_2\) is
\[
        \widehat y_{12}
        =
        2^{-4}\sum_{x_1,\ldots,x_4=\pm1}
        y(x_1,\ldots,x_4)x_1x_2 .
\]
Since \(y\) depends only on \(S\), average \(x_1x_2\) at fixed \(S\).  For \(R=4\),
\[
        \left\langle x_1x_2\right\rangle_S
        =
        \frac{S^2-4}{4\cdot3}.
\]
This follows from
\[
        S^2=4+2\sum_{a<b}x_ax_b.
\]
At fixed \(S\), all six pairs are equivalent, so each pair has average
\[
        \frac{1}{6}\left\langle\sum_{a<b}x_ax_b\right\rangle_S
        =
        \frac{S^2-4}{12}.
\]
Thus,
\[
        \widehat y_{12}
        =
        \sum_S2^{-4}\binom{4}{(4+S)/2}
        \ln\cosh(\eta S)\frac{S^2-4}{12}.
\]
Fold \(S\leftrightarrow -S\) to obtain
\[
\begin{array}{c|ccc}
|S|&0&2&4\\
\hline
W_S&6&8&2\\
(S^2-4)/12&-1/3&0&1\\
y_S&0&\ln\cosh(2\eta)&\ln\cosh(4\eta)\,\,,
\end{array}
\]

\[
        \widehat y_{12}
        =
        \frac{2}{16}\ln\cosh(4\eta)
        =
        \frac18\ln\cosh(4\eta) \,,
\]

\begin{equation}
K_{\rm MRP}^\Ising(\eta)
=
\frac18\ln\cosh(4\eta) \,.
\label{eq:isingfw}
\end{equation}

Equivalently, because \(y\) is even and replica symmetric, it has the exact \(R=4\) Walsh form is,
\[
        y=A+K_2\sum_{a<b}x_ax_b+K_4x_1x_2x_3x_4 \,.
\]
This second derivation is useful as a check.  For four replicas, even replica-symmetric Walsh functions have only three folded sectors; a constant, a two-replica pair sum, and the four-replica product.  Solving the three-sector linear system must reproduce the direct coefficient above.
Let
\[
        a=\ln\cosh(2\eta),\qquad b=\ln\cosh(4\eta).
\]
Using the three folded sectors
\[
\begin{array}{c|ccc}
|S|&0&2&4\\
\hline
\sum_{a<b}x_ax_b&-2&0&6\\
x_1x_2x_3x_4&1&-1&1\\
y&0&a&b
\end{array}
\]
gives
\[
        0=A-2K_2+K_4,\qquad
        a=A-K_4,\qquad
        b=A+6K_2+K_4.
\]
Solving,
\[
        K_2=\frac b8=\frac18\ln\cosh(4\eta),
        \qquad
        K_4=\frac{b-4a}{8}.
\]
So the pair Walsh coefficient and the previous Ising projection slope are identical.

\section{Potts models}
\label{Supp:PottsModels}

For the $q$ array Potts model the Hamiltonian is given by $-\sum_{ij}\delta_{\sigma_{i},\sigma_{j}}$, We measure the bond difference $\Delta=\sigma_{i}-\sigma_{j}$. For a \(q\)-state Potts bond, take
\[
        \Delta_a\in\Z_q,\qquad m\in\Z_q.
\]
where $m$ is the corresponding measurement outcome. The \(q\)-ary symmetric Nishimori likelihood can be written as
\[
        P(m|\Delta_a)
        =
        \frac{\ee^{J_0\delta_{m,\Delta_a}}}{\ee^{J_0}+q-1}.
\]
Equivalently, if \(\gamma\) is the probability of measuring the bond difference correctly and other \(q-1\) outcomes are equally likely, then similar to Ising model we have,
\[
        J_0
        =
        \ln\frac{1+(q-1)\gamma}{1-\gamma}.
\]
The replicated weight is
\[
        W_R^\Potts(\Delta)
        \propto
        \sum_{m=0}^{q-1}
        \prod_{a=1}^R
        \ee^{J_0\delta_{m,\Delta_a}}
        =
        \sum_{m=0}^{q-1}
        \ee^{J_0\sum_a\delta_{m,\Delta_a}}.
\]
Let
\[
        N_m(\Delta)=\#\{a:\Delta_a=m\}
        =
        \sum_{a=1}^R\delta_{m,\Delta_a}.
\]
Then
\begin{equation}
        y_R^{\rm Potts}(\Delta)
        =
        \ln\sum_{m=0}^{q-1}\ee^{J_0N_m(\Delta)} .
\label{eq:replicatedpottsweight}
\end{equation}
Potts calculation depends on the groups configurations only by the occupation vector \((N_0,\ldots,N_{q-1})\).

\subsubsection*{Potts \(q=3\)}

For Potts variables the local measurement value can be any of the \(q\) colors.  If a trial measurement value \(m\) agrees with \(N_m\) replicas, the replicated Boltzmann factor contributes \(\ee^{J_0N_m}\).  Summing over the unknown measurement color gives the exact log-weight below.  The function depends only on the occupation numbers \((N_0,\ldots,N_{q-1})\), which is why the full Fourier sum can be compressed into a small table of occupation types.

For \(q\)-state Potts variables, the \(R=4\) exact local log-weight is
\[
        y(\Delta)=
        \ln\sum_{m=0}^{q-1}\ee^{J_0N_m(\Delta)},
\]
where \(N_m(\Delta)\) is the number of replicas with \(\Delta_a=m\).
Let
\[
        y=\ee^{J_0}.
\]

For \(q=3\), the four occupation types are
\[
\begin{array}{c|c|c}
\text{type} & y_{\text{type}} & C_{\text{type}}
\\
\hline
A=(4,0,0) & \ln(y^4+2) & 3\\
B=(3,1,0) & \ln(y^3+y+1) & 6\\
C=(2,2,0) & \ln(2y^2+1) & 0\\
D=(2,1,1) & \ln(y^2+2y) & -9 .
\end{array}
\]
Here,
\[
        C_{\text{type}}
        =
        \sum_{\Delta\text{ in type}}
        \exp\left[-\frac{2\pi\ii}{3}(\Delta_1-\Delta_2)\right]
\]
is the exact character sum over all configurations in that occupation type.
Because \(y(\Delta)\) is constant inside each occupation type, the Fourier coefficient is just a weighted sum of the table entries \(y_{\text{type}}\), with weights \(C_{\text{type}}\).  This is just the finite Fourier transform after symmetry grouping.

\subsubsection*{Fourier coefficients}

The pair character coefficient is
\[
        \widehat y_{12}^{(1)}
        =
        3^{-4}\left(3y_A+6y_B-9y_D\right).
\]
Since \(K^\Potts=3\widehat y_{12}^{(1)}\),
\[
        K_{\FW}^{q=3}
        =
        \frac{3y_A+6y_B-9y_D}{27}
        =
        \frac{y_A+2y_B-3y_D}{9}.
\]
Therefore
\begin{equation}
K_{\FW}^{q=3}(y)
=
\frac19
\ln
\frac{(y^4+2)(y^3+y+1)^2}
{y^3(y+2)^3}.
\label{eq:potts3fw}
\end{equation}

The threshold equation gives
\begin{equation}
\begin{aligned}
    K_{\rm FW}^{q=3}(J_0^\ast) &= \ln(1+\sqrt{3}), \\
    J_0^\ast = 2.38365,\quad
    \gamma_c &= 0.76644,\quad
    p_c = \tfrac{2}{3}(1-\gamma_c) = 0.15571.
\end{aligned}
\end{equation}

\paragraph{Potts \(q=4\)}

The \(q=4\) Potts calculation is the same projection, but there are five occupation patterns for four replicas.  Again \(C_{\text{type}}\) is the exact sum of the pair character over all configurations of that type, while \(y_{\text{type}}\) is the common log-weight of the type.

For \(q=4\), the occupation types and character sums are
\[
\begin{array}{c|c|c}
\text{type} & y_{\text{type}} & C_{\text{type}}\\
\hline
A=(4,0,0,0) & \ln(y^4+3) & 4\\
B=(3,1,0,0) & \ln(y^3+y+2) & 16\\
C=(2,2,0,0) & \ln(2y^2+2) & 4\\
D=(2,1,1,0) & \ln(y^2+2y+1) & -16\\
E=(1,1,1,1) & \ln(4y) & -8 .
\end{array}
\]
The Fourier coefficient is
\[
        \widehat y_{12}^{(1)}
        =
        4^{-4}
        \left(4y_A+16y_B+4y_C-16y_D-8y_E\right).
\]
Since \(K^\Potts=4\widehat y_{12}^{(1)}\),
\[
        K_{\FW}^{q=4}
        =
        \frac{y_A+4y_B+y_C-4y_D-2y_E}{16}.
\]
Substituting the \(y\)-expressions,
\[
K_{\FW}^{q=4}
=
\frac1{16}
\ln
\frac{(y^4+3)(y^3+y+2)^4(2y^2+2)}
{(y^2+2y+1)^4(4y)^2}.
\]

Equivalently,
\begin{equation}
K_{\FW}^{q=4}(y)
=
\frac1{16}
\ln
\frac{(y^4+3)(y^2-y+2)^4(y^2+1)}
{8y^2(y+1)^4}.
\label{eq:potts4fw}
\end{equation}
The threshold equation is
$K_{\rm FW}^{q=4}(J_0^\ast) = \ln 3$, giving
\begin{equation}
    J_0^\ast = 2.529,\quad
    \gamma_c = 0.743,\quad
    p_c = 0.1930.
\end{equation}

\section{$\mathbb{Z}_q$ clock models}
\label{Supp:ZqClockModels}

The $q$ state Clock model is given by the Hamiltonian $H=-\sum_{ij}\cos{(\frac{2\pi}{q}(\sigma_{i}-\sigma_{j})}$. We again measure the bond difference.
For the \(q\)-state clock model, the bond difference and measurement/noise value are
\[
        \Delta_a,m\in\Z_q .
\]
The discrete von-Mises Nishimori likelihood is
\[
        P(m|\Delta_a)
        =
        \frac{
        \exp\!\left[
        J\cos\frac{2\pi(m-\Delta_a)}{q}
        \right]}
        {Z_q(J)},\
        Z_q(J)=
        \sum_{r=0}^{q-1}
        \exp\!\left[J\cos\frac{2\pi r}{q}\right].
\]
The normalization \(Z_q(J)^{-R}\) again gives only an additive constant in \(y_R\).  The nontrivial replicated weight is,
\[
        W_R^\clockmodel(\Delta)
        \propto
        \sum_{m=0}^{q-1}
        \prod_{a=1}^R
        \exp\!\left[
        J\cos\frac{2\pi(m-\Delta_a)}{q}
        \right],
\]
or
\begin{equation}
        y_R^\clockmodel(\Delta)
        =
        \ln\sum_{m=0}^{q-1}
        \exp\left[
        J\sum_{a=1}^R
        \cos\frac{2\pi(m-\Delta_a)}{q}
        \right].
\label{eq:replicatedclockweight}
\end{equation}

\subsubsection*{Clock model, general Fourier Walsh}

The clock model differs from Potts because the pair interaction resolves angular distance, not only equality or inequality.  The natural pair channel is therefore the first cosine harmonic.  Higher harmonics could also be projected, but the clean clock Hamiltonian used here has first-harmonic coupling, so the matching to \(\beta_c^\clean\) uses \(K_1^\clockmodel\).

For clock variables,
\[
        \Delta_a\in\Z_q,
\]
and the exact replicated log-weight is
\begin{equation}
        y(\Delta)
        =
        \ln\sum_{m=0}^{q-1}
        \exp\left[
        J\sum_{a=1}^4
        \cos\frac{2\pi(m-\Delta_a)}{q}
        \right].
\label{eq:clockweight}
\end{equation}
For the first harmonic \(h=1\),
\[
        X_1(\Delta)
        =
        \sum_{a<b}\cos\frac{2\pi(\Delta_a-\Delta_b)}{q}.
\]
For \(q>2\),
\[
        \langle X_1\rangle=0,\qquad
        \langle X_1^2\rangle=3,
\]
so the first-harmonic Fourier coefficient is written as
\begin{equation}
        K_1^\clockmodel(J)
        =
        2\Ree\,\widehat y_{12}^{(1)}
        =
        \frac{\langle yX_1\rangle}{\langle X_1^2\rangle}
        =
        \frac13\langle yX_1\rangle.
\label{eq:clockfw}
\end{equation}
The value \(\langle X_1^2\rangle=3\) is the uniform norm of the six-pair first-harmonic channel for \(q>2\).
For \(q=2\), the first harmonic is self-conjugate and \(\langle X_1^2\rangle=6\), so
\[
        K_1^\clockmodel(J)=\frac16\langle yX_1\rangle.
\]

Because Eq.~\eqref{eq:clockweight} is invariant under a global shift of all \(\Delta_a\), set
\[
        \Delta_1=0,\qquad
        \Delta_2=a,\qquad
        \Delta_3=b,\qquad
        \Delta_4=c.
\]
Then, for \(q>2\),
\begin{widetext}
\begin{equation}
    K_1^{\rm clock}(J)
    = \frac{1}{3q^3}
    \sum_{a,b,c=0}^{q-1}
    X_1(0,a,b,c)
    \ln\!\left[
    \sum_{m=0}^{q-1}
    \exp\!\left(
    J\sum_{n\in\{0,a,b,c\}}
    \cos\frac{2\pi(m-n)}{q}
    \right)
    \right].
\label{eq:translationreduced}
\end{equation}
\end{widetext}
For \(q=2\), replace the denominator \(3q^3\) by \(6q^3\).
This translation reduction removes one redundant sum: every orbit under the common shift of all four replica angles contributes \(q\) identical terms.

\subsection{Clock \(q=2\)}

The \(q=2\) clock model is exactly Ising in disguise.  The two clock angles are \(0\) and \(\pi\), so the cosine becomes a sign.  This is a useful check that the generalized clock formula reduces to the ordinary Walsh result.

For \(q=2\), \(\cos[\pi(m-\Delta_a)]=(-1)^m(-1)^{\Delta_a}\).  With \(x_a=(-1)^{\Delta_a}\),

\begin{widetext}
\begin{equation}
    y(\Delta)
    = \ln\sum_{m=0}^{1} \exp\!\left[J(-1)^m\sum_a x_a\right]
    = \ln\!\left[2\cosh(JS)\right],
    \qquad S = \sum_a x_a.
\end{equation}
\end{widetext}

The additive \(\ln2\) has zero pair Fourier coefficient.  Therefore the calculation is identical to the Ising Walsh calculation:
\begin{equation}
K_{q=2}^{\rm clock}(J)
=
\frac18\ln\cosh(4J).
\label{eq:clock2}
\end{equation}
Matching to \(\beta_c=\frac12\ln(1+\sqrt2)\) gives
\[
        J^\ast=1.054606205,\qquad
        T^\ast=1/J^\ast=0.948221237.
\]

\subsection{Clock \(q=3\)}
For \(q=3\), the translation-reduced sum has \(3^3=27\) configurations.  Grouping them by identical logarithms leaves only three groups with nonzero total \(X_1\) coefficient.  The table below is therefore the entire finite Fourier transform, just written compactly.

For \(q=3\), Eq.~\eqref{eq:translationreduced} has denominator
\[
        3q^3=81.
\]
Only three grouped logarithms have nonzero total \(X_1\) coefficient:
\[
\begin{array}{c|c}
C_g&L_g(J)\\
\hline
6&\ee^{4J}+2\ee^{-2J}\\
12&\ee^{5J/2}+\ee^{-J/2}+\ee^{-2J}\\
-18&\ee^{J}+2\ee^{-J/2}.
\end{array}
\]
Therefore
\begin{equation}
K_{q=3}^\clockmodel(J)
=
\frac1{81}
\left[
6\ln L_1+12\ln L_2-18\ln L_3
\right].
\label{eq:clock3}
\end{equation}
the clean critical coupling from the supplied formula is
\[
        \beta_{sd}(q=3)=\frac23\ln(1+\sqrt3)=0.670035026.
\]
Solving \(K_{q=3}^\clockmodel(J^\ast)=\beta_{sd}\) gives
\[
        J^\ast=1.589102389,\qquad
        T^\ast=0.629286072.
\]

\subsection{Clock \(q=4\)}

For \(q=4\), the grouped logarithms contain \(\cosh J\), \(\cosh 2J\), and constants, but the projection coefficients cancel everything except \(\ln\cosh(2J)\).

For \(q=4\), Eq.~\eqref{eq:translationreduced} has denominator \(3q^3=192\).  Grouping the \(4^3\) translation-reduced configurations by their \(m\)-sum \(L_g(J)\) and coefficient \(C_g=\sum X_1\) gives
\[
        K_{q=4}^{\rm clock}(J)=\frac1{192}\sum_g C_g\ln L_g(J),
\]
with the nonzero groups
\[
\begin{array}{c|c}
C_g&L_g(J)\\
\hline
6&2+2\cosh(4J)\\
24&2\cosh(3J)+2\cosh J\\
12&4\cosh(2J)\\
-24&4\cosh J\\
-18&4 .
\end{array}
\]
Now use
\[
        2+2\cosh4J=4\cosh^2 2J,\
        2\cosh3J+2\cosh J=4\cosh2J\cosh J.
\]
Then
\begin{align}
192K_{q=4}^{\rm clock}
&=
6\ln(4\cosh^2 2J)
 +24\ln(4\cosh2J\cosh J)\\
&\quad
 +12\ln(4\cosh2J)
 -24\ln(4\cosh J)
 -18\ln4 .
\end{align}

\begin{equation}
K_{q=4}^{\rm clock}(J)=\frac14\ln\cosh(2J).
\label{eq:clock4}
\end{equation}
Matching to the clean four-state clock coupling \(\beta_c=\ln(1+\sqrt2)=0.881373587\) gives
\[
        J^\ast=2.109212410,\qquad
        T^\ast=0.474110618.
\]

\subsection{Clock \(q=5\)}

For \(q=5\), the exact grouped expression is already compact but no longer collapses to a single elementary logarithm.  The representative triples \((a,b,c)\) below stand for translation-reduced replica configurations \((0,a,b,c)\).  The coefficients \(C_r\) include all permutations and symmetry-related configurations with the same logarithm.

For \(q=5\), Eq.~\eqref{eq:translationreduced} has denominator
\[
        3q^3=375.
\]
It is shortest to keep the exact clock character sum in representative form.  Define
\[
        L_{abc}^{(5)}(J)
        =
        \sum_{m=0}^{4}
        \exp\left[
        J\sum_{n\in\{0,a,b,c\}}
        \cos\frac{2\pi(m-n)}{5}
        \right].
\]
Grouping translation-reduced configurations with identical \(L_{abc}^{(5)}\), the nonzero coefficients are
\[
\begin{array}{c|c}
C_r&(a,b,c)\\
\hline
18+6\sqrt5&(0,0,1)\\
6+6\sqrt5&(0,1,1)\\
-3+9\sqrt5&(0,1,4)\\
-6+6\sqrt5&(0,1,2)\\
6&(0,0,0)\\
18-6\sqrt5&(0,0,2)\\
6-6\sqrt5&(0,2,2)\\
-6-6\sqrt5&(0,1,3)\\
-3-9\sqrt5&(0,2,3)\\
-36&(1,2,3).
\end{array}
\]
Thus
\begin{equation}
K_{q=5}^\clockmodel(J)
=\frac1{375}\sum_{r=1}^{10}C_r\ln L_r(J),
\qquad
L_r(J)=L_{a_rb_rc_r}^{(5)}(J).
\label{eq:clock5}
\end{equation}

The clean BKT temperatures supplied in the table are,
\[
        T_{c1}^{\rm clean} = 0.9521(2),
        \qquad
        T_{c2}^{\rm clean} = 0.9059(2),
\]
so the clean inverse-temperature inputs for the Nishimori projection are
\[
        \beta_{c1}^{\rm clean} = 1.050309841,
        \qquad
        \beta_{c2}^{\rm clean} = 1.103874600.
\]
Solving $K_{q=5}^{\clockmodel}(J^\ast) = \beta_c^{\rm clean}$ gives

\begin{center}
\begin{tabular}{lccc}
clean input & $\beta_c^{\rm clean}$ & $J^\ast$ & $T^\ast = 1/J^\ast$ \\ \hline
$T_{c1}^{\rm clean} = 0.9521$ & 1.0503098 & 2.4993839 & $T_1^\ast = 0.4000985$ \\
$T_{c2}^{\rm clean} = 0.9059$ & 1.1038746 & 2.6110358 & $T_2^\ast = 0.3829897$ \\
\end{tabular}
\end{center}

\subsection{Clock \(q=6\)}

For \(q=6\), the same finite character sum produces eight logarithms with nonzero first-harmonic projection.  This is still an exact calculation: all \(6^3\) translation-reduced configurations are accounted for by the grouped coefficients.

For \(q=6\), Eq.~\eqref{eq:translationreduced} has denominator
\[
        3q^3=3\cdot6^3=648.
\]
The exact finite Fourier sum groups into eight nonzero logarithms.  A ninth group has zero total \(X_1\) coefficient and drops out.  Write
\[
        K_{q=6}^\clockmodel(J)
        =
        \frac1{648}\sum_{r=1}^{8} C_r\ln L_r(J).
\]
The grouped terms are
\[
\begin{array}{c|l}
C_r&L_r(J)\\
\hline
6&
\ee^{4J}+2\ee^{2J}+2\ee^{-2J}+\ee^{-4J}\\
36&
\ee^{7J/2}+\ee^{5J/2}+\ee^J+\ee^{-J}+\ee^{-5J/2}+\ee^{-7J/2}\\
30&
\ee^{3J}+2\ee^{3J/2}+2\ee^{-3J/2}+\ee^{-3J}\\
24&
2\ee^{3J}+2+2\ee^{-3J}\\
48&
\ee^{5J/2}+\ee^{2J}+\ee^{J/2}+\ee^{-J/2}+\ee^{-2J}+\ee^{-5J/2}\\
-24&
2\ee^{3J/2}+2+2\ee^{-3J/2}\\
-90&
\ee^J+2\ee^{J/2}+2\ee^{-J/2}+\ee^{-J}\\
-30&
6 .
\end{array}
\]
Thus
\begin{equation}
\begin{aligned}
K_{q=6}^\clockmodel(J)
&=\frac1{648}\Big[
6\ln L_1+36\ln L_2+30\ln L_3+24\ln L_4
\\
&\qquad\qquad
+48\ln L_5-24\ln L_6-90\ln L_7-30\ln6
\Big].
\end{aligned}
\label{eq:clock6}
\end{equation}
This is the direct Fourier-character formula for the \(q=6\) first-harmonic clock coupling.

The clean BKT temperatures supplied in the table are,
\[
        T_{c1}^{\rm clean} = 0.9127(5),
        \qquad
        T_{c2}^{\rm clean} = 0.6901(4).
\]
Therefore
\[
        \beta_{c1}^{\rm clean} = 1/T_{c1}^{\rm clean} = 1.095650,\
        \beta_{c2}^{\rm clean} = 1/T_{c2}^{\rm clean} = 1.449065.
\]
Solving $K_{q=6}^{\clockmodel}(J^\ast) = \beta_c^{\rm clean}$ gives

\begin{center}
\begin{tabular}{lccc}
clean input & $\beta_c^{\rm clean}$ & $J^\ast$ & $T^\ast = 1/J^\ast$ \\ \hline
$T_{c1}^{\rm clean} = 0.9127$ & 1.09565 & 2.58333 & $T_1^\ast = 0.387096040$ \\
$T_{c2}^{\rm clean} = 0.6901$ & 1.44906 & 3.30997 & $T_2^\ast = 0.302116929$ \\
\end{tabular}
\end{center}

As in the Ising case the Nishimori temperature can be written in terms of measurement strength ($\gamma_c$) with these change of variables,

\begin{equation}
\gamma_c
=\Big\langle \cos\frac{2\pi(m-\Delta)}{q}\Big\rangle_{J^\ast}
=\frac{\sum_{k=0}^{q-1}\cos\!\big(\frac{2\pi k}{q}\big)\,
        e^{J^\ast\cos(2\pi k/q)}}
      {\sum_{k=0}^{q-1} e^{J^\ast\cos(2\pi k/q)}}\,,
\end{equation}


\subsection{$\mathbb{Z}_q$ clock models for large $q$}

For the high-temperature BKT branch, using the \(q=8\) results from Ref.~\cite{PhysRevE.101.060105},
\[
      \beta_{c1}=1.10375,
\]
as the finite-\(q\) proxy for the continuum clean value, the solution of \(K_\infty(J_{\mathrm{hi}}^\ast)=\beta_{c1}^{\clean}\) is
\[
T_{\mathrm{1}}^\ast(\infty)\simeq0.385079
\]

For the lower BKT transition, the Jose-Kadanoff-Kirkpatrick-Nelson (JKKN) asymptotic clean coupling is
\[
        \beta_{c2}(q)\sim \frac{q^2}{8\pi}.
\]
At large coupling,
\[
        \ln I_0(J\rho)=J\rho+O(\ln J),
\]
and therefore
\[
        K_\infty(J)=c_\infty J+O(\ln J),
        \qquad
        c_\infty=
        \frac{1}{3(2\pi)^3}
        \int X(\theta)\rho(\theta)\,d^3\theta .
\]
Numerical quadrature gives
\[
        c_\infty=0.487385,
        \qquad
        8\pi c_\infty=12.249325.
\]

\begin{equation}
J_2^\ast(q)
        \sim
        \frac{q^2}{8\pi c_\infty}
        =
        \frac{q^2}{12.249325},
        \qquad
        T_2^\ast(q)
        \sim
        \frac{12.249325}{q^2}.
\label{eq:largeqlower}
\end{equation}

\section{Uncertainty propagation}
\label{Supp:Uncertainty}


Here we calculate the uncertainty propagated to our projection threshold from the uncertainty of the clean critical temperature. We use
the same first-order independent-variable rule used by Julia's common \textit{Measurements.jl} package.  The notation \(x(y)\) means \(x\pm y\) in the last displayed digits, then
\[
        \sigma_u=\left|\frac{\partial f}{\partial \beta}\right|\sigma_\beta .
\]
For direct Ising maps this can be applied directly to
\[
        \eta(\beta)=\frac14\arcosh\!\left(\ee^{8\beta}\right),
        \qquad
        p(\beta)=\frac{1-\tanh\eta(\beta)}{2}.
\]
The derivative used for the Ising error bars is
\[
        \frac{d\eta}{d\beta}
        =
        \frac{2\ee^{8\beta}}{\sqrt{\ee^{16\beta}-1}},
        \qquad
        \frac{dp}{d\beta}
        =
        -\frac12\operatorname{sech}^2\eta\,
        \frac{d\eta}{d\beta}.
\]
For implicit clock or Potts equations
\[
        K(J^\ast)=\beta,
\]
differentiate implicitly:
\[
        K'(J^\ast)\,\mathrm dJ^\ast=\mathrm d\beta,
        \qquad
        \sigma_{J^\ast}=\frac{\sigma_\beta}{|K'(J^\ast)|},
        \qquad
        \sigma_T=\frac{\sigma_{J^\ast}}{(J^\ast)^2}.
\]
For a grouped formula \(K(J)=D^{-1}\sum_r C_r\ln L_r(J)\), the derivative is computed from the same grouped data:
\[
        K'(J)=\frac1D\sum_r C_r\frac{L_r'(J)}{L_r(J)}.
\]
For an ungrouped finite clock sum, this is equivalently obtained by differentiating Eq.~\eqref{eq:translationreduced} term by term. No numerical finite-difference derivative is needed. The uncertainties calculated from this method is used in Table~\ref{tab:thresholds}

\begin{table}[t]
\caption{{\bf Clock thresholds} inserted into the Gilbert--Varshamov
self-dual entropy condition.}
\label{tab:clock-gv}
\centering
\scriptsize
\setlength{\tabcolsep}{4pt}
\begin{ruledtabular}
\begin{tabular}{clll}
\(q\) & \multicolumn{1}{c}{\(T_1^\ast\)} & \multicolumn{1}{c}{\(T_2^\ast\)} & \multicolumn{1}{c}{\(H_{\rm GV}/\ln q\)} \\
2 & 0.948221 & -- & 0.988953 \\
3 & 0.629286 & -- & 0.983814 \\
4 & 0.474111 & -- & 0.988954 \\
5 & 0.400099(74) & 0.382990(75) & 1.018125(138) \\
6 & 0.387096(184) & 0.302117(158) & 1.046444(301) \\
7 & 0.385290(184) & 0.237933(125) & 1.054064(258) \\
8 & 0.385060(184) & 0.190686(129) & 1.055923(269) \\
\end{tabular}
\end{ruledtabular}
\end{table}

\begin{table}[t]
\caption{{\bf Large-\(q\) entropy sums} for the two projected clock transitions.}
\label{tab:clock-large-q-entropy}
\centering
\scriptsize
\setlength{\tabcolsep}{3.5pt}
\begin{ruledtabular}
\begin{tabular}{crrrr}
\(q\) &
\(H_q(T_1^\ast)\) &
\(H_q(T_2^\ast)\) &
\(H_{\rm sum}-\ln q\) &
\(H_{\rm sum}/\ln q\) \\
10 & 1.555082 & 0.843854 & 0.096351 & 1.041845(141) \\
12 & 1.737421 & 0.833056 & 0.085570 & 1.034436(131) \\
20 & 2.248247 & 0.818845 & 0.071360 & 1.023820(108) \\
50 & 3.164538 & 0.812629 & 0.065144 & 1.016652(83) \\
100 & 3.857685 & 0.811763 & 0.064278 & 1.013958(70) \\
1000 & 6.160270 & 0.811479 & 0.063994 & 1.009264(47) \\
\end{tabular}
\end{ruledtabular}
\end{table}

\section{Variance Decomposition}
\label{sec:varianceprojection}
The single channel calculation keeps only the leading pair channel.  It is therefore useful to ask what part of the exact replicated log-weight is being discarded. For the clock model define the harmonic pair observables
\[
        X_h(\Delta)
        =
        \sum_{a<b}
        \cos\!\left[
        \frac{2\pi h}{q}(\Delta_a-\Delta_b)
        \right],
        \qquad
        h=1,\ldots,\lfloor q/2\rfloor .
\]
The threshold projection used above is the \(h=1\) coefficient.  For a sector decomposition with multiplicities \(W_s\), sector log-weights \(y_s\), and sector observables \(X_{h,s}\), the grouped Fourier--Walsh inner product is
\[
        \langle f,g\rangle_{\FW}
        =
        \frac{1}{\mathcal N}
        \sum_s W_s f_s g_s,
        \qquad
        \mathcal N=\sum_s W_s=q^4 .
\]
This is exactly the uniform finite-group inner product, only written after grouping equivalent configurations.  Define the centered functions,
\[
        \widetilde y
        =
        y-\langle y,1\rangle_{\FW},
        \qquad
        \widetilde X_h
        =
        X_h-\langle X_h,1\rangle_{\FW}.
\]
The Fourier--Walsh projection of \(\widetilde y\) onto the \(h\)-th clock-pair harmonic is
\[
        \Pi_h\widetilde y
        =
        K_h\widetilde X_h,
        \qquad
        K_h
        =
        \frac{\langle \widetilde y,\widetilde X_h\rangle_{\FW}}
        {\langle \widetilde X_h,\widetilde X_h\rangle_{\FW}} .
\]
The Fourier--Walsh norm fraction carried by harmonic \(h\) is
\[
        \mathcal V_h
        =
        \|\Pi_h\widetilde y\|_{\FW}^2
        =
        K_h^2\|\widetilde X_h\|_{\FW}^2 .
\]
The total centered log-weight norm is
\[
        \mathcal V_{\mathrm{tot}}
        =
        \|\widetilde y\|_{\FW}^2 .
\]
Because distinct finite-group harmonics are orthogonal under \(\langle\cdot,\cdot\rangle_{\FW}\), the residual Fourier--Walsh norm after removing all single-copy clock harmonics is
\[
        \mathcal V_{\mathrm{rep}}
        =
        \left\|
        \widetilde y-\sum_{h=1}^{\lfloor q/2\rfloor}\Pi_h\widetilde y
        \right\|_{\FW}^2
        =
        \mathcal V_{\mathrm{tot}}
        -
        \sum_{h=1}^{\lfloor q/2\rfloor}\mathcal V_h ,
\]
where the last term is the purely replica-sector residual, orthogonal to every single-copy clock harmonic included in the fit.

\begin{table}[t]
\caption{{\bf Clock harmonic Fourier--Walsh diagnostic} at the projected
transition points.  The last column gives the norm fraction carried by
the first harmonic alone.}
\label{tab:harmonic-ratios}
\centering
\scriptsize
\setlength{\tabcolsep}{3.5pt}
\begin{ruledtabular}
\begin{tabular}{lccc}
Case & \(K_1\) & \((K_2,K_3,K_4)/K_1\) &
\(\mathcal V_1/\mathcal V_{\rm tot}\) \\
\(q=2,\ \beta_c=0.4407\)
& 0.4407 & \((--,--,--)\) & 0.9392 \\
\(q=4,\ \beta_c=0.8814\)
& 0.8814 & \((0.0000,--,--)\) & 0.9392 \\
\(q=6,\ \beta_{c1}=1.11\)
& 1.1100 & \((-0.0353,+0.0087,--)\) & 0.9577 \\
\(q=6,\ \beta_{c2}=1.42\)
& 1.4200 & \((-0.0378,+0.0127,--)\) & 0.9507 \\
\(q=8,\ \beta_{c1}=1.1115\)
& 1.1115 & \((-0.0371,+0.0051,-0.0010)\) & 0.9602 \\
\(q=8,\ \beta_{c2}=2.36\)
& 2.3600 & \((-0.0446,+0.0126,-0.0018)\) & 0.9450 \\
\end{tabular}
\end{ruledtabular}
\end{table}

The important point is that the first harmonic carries about \(94\%-96\%\) (See Table~\ref{tab:harmonic-ratios}) of the weighted variance at these transition points.  The higher clock harmonics are tiny, in the displayed cases their total contribution is at most \(0.20\%\) of the variance.  The remaining few percent is not another clock coupling.  It lies in directions orthogonal to the full clock-harmonic subspace, so it has no direct single-copy clean-clock counterpart.

\begin{table}[t]
\caption{{\bf Decomposition of the projected clock weight} at the transition
points.  Entries are percentages of the total Walsh norm.}
\label{tab:clock-weight-decomposition}
\centering
\scriptsize
\setlength{\tabcolsep}{3.5pt}
\begin{ruledtabular}
\begin{tabular}{lccc}
Case & \(V_1\) & \(V_{k>1}\) & residual \\
\(q=2,\ \beta_c=0.4407\)
& 93.92 & 0.00 & 6.08 \\
\(q=4,\ \beta_c=0.8814\)
& 93.92 & 0.00 & 6.08 \\
\(q=6,\ \beta_{c1}=1.11\)
& 95.77 & 0.13 & 4.10 \\
\(\phantom{q=6,\ }\beta_{c2}=1.42\)
& 95.07 & 0.17 & 4.76 \\
\(q=8,\ \beta_{c1}=1.1115\)
& 96.02 & 0.14 & 3.85 \\
\(\phantom{q=8,\ }\beta_{c2}=2.36\)
& 94.50 & 0.20 & 5.30 \\
\end{tabular}
\end{ruledtabular}
\end{table}

\section{Bethe lattice}
\label{Supp:Bethe}
On the Bethe lattice the object our
construction reproduces is the linearized cavity instability, a
strictly two-replica quantity. The complete local algebra for this
instability is therefore $R=2$, consistent with the minimality
discussion of the main text. The $R=2$
pair-overlap Walsh coefficient is the exact projection coordinate.

\paragraph{Ising.}
For one bond $\langle ij\rangle$, define
\[
    x_a=\sigma_i^{(a)}\sigma_j^{(a)}=\pm1,
    \qquad a=1,2,
\]
Summing
the shared measurement outcome, the two-replica bond weight and
log-weight are
\[
\begin{aligned}
    W_2(x_1,x_2)
    &=2\cosh\!\left[\eta(x_1+x_2)\right],\\
    y_2(x_1,x_2)
    &=\log\cosh\!\left[\eta(x_1+x_2)\right].
\end{aligned}
\]
The Walsh basis on $(\mathbb Z_2)^2$ is $\{1,x_1,x_2,x_1x_2\}$, so
\[
    y_2(x_1,x_2)=A+B_1x_1+B_2x_2+\kappa_2\, x_1x_2\,.
\]
Since $y_2$ is invariant under the exchange $x_1\leftrightarrow x_2$
and under the simultaneous flip $(x_1,x_2)\mapsto(-x_1,-x_2)$, the
linear coefficients vanish, $B_1=B_2=0$, and
\[
    A=\kappa_2(\eta)=\tfrac12\log\cosh(2\eta),
    \qquad
    y_2=\kappa_2(\eta)\left(1+x_1x_2\right).
\]
The nontrivial character factorizes on sites,
\[
    x_1x_2=\bigl(\sigma_i^{(1)}\sigma_i^{(2)}\bigr)
           \bigl(\sigma_j^{(1)}\sigma_j^{(2)}\bigr)
          =Q_i\,Q_j\,,
\]
so the overlap spin $Q_i=\sigma_i^{(1)}\sigma_i^{(2)}$ sees an
ordinary Ising coupling $\kappa_2$, with the exact identity
\[
    \tanh\kappa_2(\eta)=\tanh^2\eta\,.
\]
On a Bethe lattice of coordination $z$, branching number $c=z-1$, the
clean instability is $c\tanh K_c^{\rm clean}=1$; for Ising the
transition is continuous, so this linear instability is the critical
point itself. Matching $\kappa_2(\eta_c)=K_c^{\rm clean}$ gives
\[
    \tanh^2\eta_c=\frac1c\,,
\]
and with $\tanh\eta=\gamma=1-2p$ on the Nishimori line,
\[
    p_c^{R=2}=\frac12\left(1-\frac1{\sqrt{z-1}}\right),
\]
Eq.~\eqref{eq:KS} of the main text. The projection is not an approximation
here, it isolates the pair-overlap channel whose instability,
$c\tanh^2\eta_c=1$, is the KS condition of the binary
symmetric channel.

\paragraph{Potts.}
The same steps carry over with the $q$-ary likelihood of Appendix~\ref{app:PottsModels}, 
$J_0=\ln[(1+(q-1)\gamma)/(1-\gamma)]$ and
$\gamma=(e^{J_0}-1)/(e^{J_0}+q-1)$. The two-replica bond weight takes
only two values,
\[
\begin{aligned}
W_2(\Delta_1,\Delta_2)
&=\sum_{m=0}^{q-1}
e^{J_0(\delta_{m,\Delta_1}+\delta_{m,\Delta_2})} \\
&=
\begin{cases}
e^{2J_0}+q-1\,, & \Delta_1=\Delta_2\,,\\[2pt]
2e^{J_0}+q-2\,, & \Delta_1\neq\Delta_2\,.
\end{cases}
\end{aligned}
\]
so the $R=2$ bond is again a pure Potts coupling. Projecting onto the
centered coincidence channel $x=\delta_{\Delta_1,\Delta_2}-1/q$ gives
\[
    \kappa_2(J_0)=\ln\frac{e^{2J_0}+q-1}{2e^{J_0}+q-2}\,,
\]
and with $v(K)=(e^K-1)/(e^K+q-1)$, the second transfer-matrix
eigenvalue of a clean Potts bond, the exact identity
\[
    v\!\left(\kappa_2(J_0)\right)
    =\frac{(e^{J_0}-1)^2}{(e^{J_0}+q-1)^2}
    =\gamma^2\,,
\]
the Potts analog of $\tanh\kappa_2=\tanh^2\eta$. The clean linear
instability on the Bethe lattice is $c\,v(K_{\rm lin})=1$, i.e.
$K_{\rm lin}=\ln[(z+q-2)/(z-2)]$. Matching
$\kappa_2(J_0^\ast)=K_{\rm lin}$ then yields
\[
    c\,\gamma_c^2=1\,,
    \qquad
    p_c^{R=2}=\frac{q-1}{q}\left(1-\frac1{\sqrt{z-1}}\right).
\]
This is the KS linear instability of the $q$-ary
symmetric channel. For $q=2$ it coincides with the Ising
reconstruction threshold. For $q>2$ one should
distinguish this linear instability from possible first-order or
non-Kesten--Stigum reconstruction transitions; here it is the correct
object because the comparison is explicitly to the two-replica linear
pair-channel instability.



\end{document}